\documentclass[a4paper,11pt]{article}

\usepackage{listings}             
\usepackage{csquotes}
\usepackage{xcolor}
\usepackage{comment}

\usepackage{amsmath,amsfonts,amssymb}

\def\D{\partial}
\def\vb{{\bar{v}}}

\def\e{\varepsilon}
\def\ke{k-\varepsilon}
\def\a{\alpha}
\def\pycalcr{\textbf{pyCALC-RANS}}
\def\pycalc{\textbf{pyCALC-LES}}
\def\reich{{Reichardt's law}}

\usepackage{tikz,xifthen}
\usetikzlibrary{patterns}
\usepackage{pgfplots}

\def\uv{\overline{u'v'}}

\usepackage{subcaption}

\usepackage[hyphens]{url}  

\makeatletter
\pgfarrowsdeclare{myarrow}{myarrow}{
\pgfarrowsleftextend{-5\pgflinewidth}
\pgfarrowsrightextend{5\pgflinewidth}
}
{
            \newlength\pgf@my@length
            \pgf@my@length = 1cm
            \divide \pgf@my@length by 16
            \pgfpathmoveto{\pgfpoint{-.7\pgf@my@length}{0pt}}
            \pgfpathlineto{\pgfpoint{-.7\pgf@my@length}{-.8\pgf@my@length}}
            \pgfpathlineto{\pgfpoint{1\pgf@my@length}{0pt}}
            \pgfpathlineto{\pgfpoint{-.7\pgf@my@length}{.8\pgf@my@length}}
            \pgfpathlineto{\pgfpoint{-.7\pgf@my@length}{0pt}}
            \pgfusepathqfillstroke
}
\makeatother

\def\solidline{\textcolor{blue}{\protect\rule[0.15\baselineskip]{0.6cm}{1.5pt}}\hspace*{0.1cm}}

\def\dashedline{\textcolor{red}{\protect\rule[0.15\baselineskip]{0.2cm}{1.5pt}}%
\hspace*{0.1cm}\textcolor{red}{\protect\rule[0.15\baselineskip]{0.2cm}{1.5pt}}\hspace*{0.1cm}}

\def\dashedblackline{\textcolor{black}{\protect\rule[0.15\baselineskip]{0.2cm}{2.5pt}}%
\hspace*{0.1cm}\textcolor{black}{\protect\rule[0.15\baselineskip]{0.2cm}{2.5pt}}\hspace*{0.1cm}}

\usepackage{listofitems} 
\usetikzlibrary{arrows.meta} 
\usepackage[outline]{contour} 
\contourlength{1.4pt}
\tikzset{>=latex} 
\usepackage{xcolor}
\colorlet{myred}{red!80!black}
\colorlet{myblue}{blue!80!black}
\colorlet{mygreen}{green!60!black}
\colorlet{myorange}{orange!70!red!60!black}
\colorlet{mydarkred}{red!30!black}
\colorlet{mydarkblue}{blue!40!black}
\colorlet{mydarkgreen}{green!30!black}
\tikzstyle{node}=[thick,circle,draw=myblue,minimum size=22,inner sep=0.5,outer sep=0.6]
\tikzstyle{node in}=[node,green!20!black,draw=mygreen!30!black,fill=mygreen!25]
\tikzstyle{node hidden}=[node,blue!20!black,draw=myblue!30!black,fill=myblue!20]
\tikzstyle{node convol}=[node,orange!20!black,draw=myorange!30!black,fill=myorange!20]
\tikzstyle{node out}=[node,red!20!black,draw=myred!30!black,fill=myred!20]
\tikzstyle{connect}=[thick,mydarkblue] 
\tikzstyle{connect arrow}=[-{Latex[length=4,width=3.5]},thick,mydarkblue,shorten <=0.5,shorten >=1]
\tikzset{ 
  node 1/.style={node in},
  node 2/.style={node hidden},
  node 3/.style={node out},
}
\def\nstyle{int(\lay<\Nnodlen?min(2,\lay):3)} 

\usepackage[numbers]{natbib}
\usepackage[colorlinks=true,linkcolor=blue,anchorcolor=blue,urlcolor=blue,citecolor=blue]{hyperref}
\usepackage{lipsum}
\usepackage[colorlinks=true,linkcolor=blue,anchorcolor=blue,urlcolor=blue,citecolor=blue]{hyperref}

\title{Using Physics Informed Neural Network (PINN) and Neural Network (NN) to Improve a $k-\omega$ Turbulence Model}
\author{
Lars Davidson \\ 
Div. of  Fluid Dynamics\\
Dept of  Mechanical Engineering\\
Chalmers University of Technology, Gothenburg, Sweden \\
Email: lada@chalmers.se
}
\date{}

\begin{document}
\lstset{language=Python,breaklines,
    numbers=right,
    stepnumber=1,
  basicstyle=\ttfamily,
  columns=fullflexible,
  frame=single,  
  breaklines=true,
  postbreak=\mbox{\textcolor{red}{$\hookrightarrow$}\space},
}

\newcommand{\sbb}{\bar{s}}

\maketitle

\begin{abstract}
The Wilcox $k-\omega$ turbulence model predicts turbulent boundary layers well, both fully-developed channel flows and flat-plate boundary
layers. However, it predicts too low a turbulent kinetic energy. This is a feature it shares with most other two-equation
turbulence models. When comparing the terms in the $k$ equations with DNS data it is found that the production
and dissipation terms are well predicted but the turbulent diffusion is not. In the present work the poor modeling of the turbulent
diffusion is improved using Physics Informed Neural Network (PINN) and Neural Network (NN).
The $k$ equation is turned into
an ordinary differential equation for the turbulent viscosity in the $k$ equation, $\nu_{t,PINN}$, which  is
solved using PINN. A new turbulent Prandtl number is then computed as $\sigma_{k,PINN} = \nu_{t}/\nu_{t,PINN}$ where $\nu_t = k/\omega$.
Hence, the turbulent Prandtl number, $\sigma_{k,PINN}$, is determined using PINN.
This is followed by the use of DNS data for estimating $C_{k,PINN}$ and
$C_{\omega2,PINN}$ which appear in the destruction terms in the $k$ and $\omega$ equation, respectively. 
Neural networks are then used to generalize these results
and thus construct a turbulence model.
The new turbulence model, called the
 $k-\omega$-PINN-NN model, is shown to produce excellent velocity, skin friction and turbulent kinetic   profiles
in  channel flow at $Re_\tau = 2\, 000$,  $5\, 200$ and   $Re_\tau = 10\, 000$
as well as in flat-plate boundary layer flow (slightly too large a $k$ for the latter case).
 The  $k-\omega$-PINN-NN model is also used for predicting the flow
over a periodic hill and the agreement with DNS is very good.
At the end of the Conclusions, 
we give an example on how a NN model can be replaced with a Python symbolic regression (pySR); the latter 
may conveniently be imported in  commercial CFD codes.
All Python PINN, NN and pySR
 scripts as well as the Python CFD code can be downloaded~\citep{davidson-pinn-nn:25}.
\end{abstract}


\section{Introduction}

Eddy-viscosity RANS (Reynolds-Averaged Navier-Stokes) turbulence models have been developed with the object of predicting
a correct velocity field, $\vb_i$.  The most common turbulence models are the $\ke$ and the $k-\omega$ models where $k$, $\e$ and $\omega$
denote the turbulent kinetic energy, its dissipation rate and the specific dissipation $\omega = \e/(\beta^* k)$
 ($\beta^*$ is a constant), respectively. Both
turbulence models 
include five tuning constants. These constants can be improved using PINN.

In \cite{yazdani:24} they use PINN to optimize the five constants $\a$, $\beta^*$, $\beta$, $\sigma_k$ and $\sigma_\omega$
 in the $k-\omega$ model. The constant $\beta^*$ appears in the dissipation term in the $k$ equation.  $\sigma_k$ and $\sigma_\omega$
are included in the turbulent diffusion term in the $k$ and $\omega$ equation, respectively. The constants  $\a$ and $\beta$ are
found in the production and destruction term, respectively,  in the $\omega$ equation.
 DNS data of the flow over a periodic hill at $Re = 5600$ are used.
Using these DNS data, they compute all terms in the 2D RANS equations comprising
of the two momentum 
equations ($\vb_1$ and $\vb_2$), the continuity equation, the $k$ and $\omega$ equations. The loss function is defined
as the sum of $(\hat{Q}^n_i - \tilde{Q}^n_i)^2$ where $\hat{Q}$ and   $\tilde{Q}$ denote DNS and PINN value, respectively. 
Subscript $i$ represents  $5000$ arbitrary chosen DNS data  points and superscript $n$ corresponds to  
$\vb_1$, $\vb_2$, continuity equation, $k$ or $\e$.  The residuals (square of $L_2$ norm) of 
the five governing equations,  $Q^n$, are added to the loss function.
They use PINN to
find optimal values of the five coefficient in the $k-\omega$ model. PINN gives modified values for $\a = 2.9719$ and $\sigma_\omega= 1.2685$
(baseline values are $2$ and $0.5$)  whereas the 
other three constants are not changed. Then they carry out RANS simulations of the flow over the same periodic
hill that was used for training 
and compare with RANS simulations using the standard values for the coefficients. They find that the new PINN coefficients
give somewhat better results. 

Luo et~al. \cite{luo:20} improve the five constants  $C_\mu$, $C_{\e 1}$,  $C_{\e 2}$, $\sigma_k$ and $\sigma_\e$ 
 in the $\ke$ model using PINN.  The turbulent viscosity is linearly dependent on $C_\mu$. The constants 
$C_{\e 1}$ and  $C_{\e 2}$ are part
of the production and dissipation, respectively, in the $\e$ equation.  $\sigma_k$ and $\sigma_\e$ appear in 
turbulent diffusion terms in the $k$ and $\e$ equation, respectively.
The authors define the  loss function as   $(\hat{Q}^n_i - \tilde{Q}^n_i)^2$ (see above) 
at the DNS data points
where $Q^n$ is $k$ or $\e$. Superscript $n=1$ for the $k$ equation and $n=2$ for the $\e$ equation.
 Subscript $i$ represents all DNS data points.
The residuals of the transport equations of $k$ and $\e$ --  multiplied by penalty functions --  are added to the loss function.
All terms in the $k$ and $\e$ equations are taken from DNS
of a converging-diverging channel.  New values of the constants are found by PINN ($C_\mu$ keeps its standard value of $0.09$). 
Finally, RANS simulations are carried out for the converging-diverging channel flow (the same flow that was used
when training the PINN) using the new PINN-optimized $\ke$  constants. 
Somewhat better results
are obtained compared with the standard $\ke$  constants.

Thakur et~al. \cite{hakut:24} use PINN to predict a diffusion coefficient. They study an unsteady, two dimensional
 convection-diffusion concentration equation, $c=c(x,y,t)$, 
with a spatially dependent diffusion coefficient, $D=D(x,y)$. The object is to predict $D$. The loss function, $L$,
is
\begin{eqnarray*}
L = \frac{1}{N \sigma_c} \sum_{i=1}^N \left | \tilde{c}_i - c_i\right|^2
 + \frac{1}{N^e \sigma_c} \sum_{i=1}^{N_e} \left | \tilde{c}_i - c_i^p\right|^2
\end{eqnarray*}
where $c_i$ denotes true data predicted with CFD using a prescribed (true), varying diffusion coefficient ($D_{true}$ --
which is the prescribed $D$ in the CFD simulation -- varies
linearly, $\sin$, $\tanh$ etc) and $c^p$ is obtained from the explicit unsteady diffusion equation;
in the unsteady term,  $\tilde{c}$ is used as the old time  value of $c^p$.
 $\sigma_c$ is the standard deviation of the concentration. $D$ and $\tilde{c}$ are obtained from two neural networks.

\begin{figure}
\centering
\begin{subfigure}[t]{0.5\textwidth}
\centering
\includegraphics[scale=0.28,clip=]{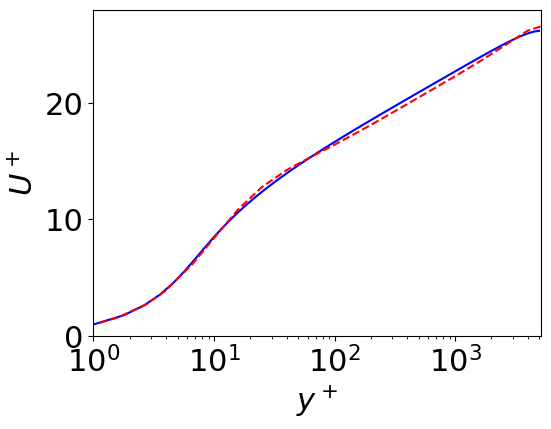}
\caption{Mean velocity.}
\end{subfigure}%
\begin{subfigure}[t]{0.5\textwidth}
\centering
\includegraphics[scale=0.28,clip=]{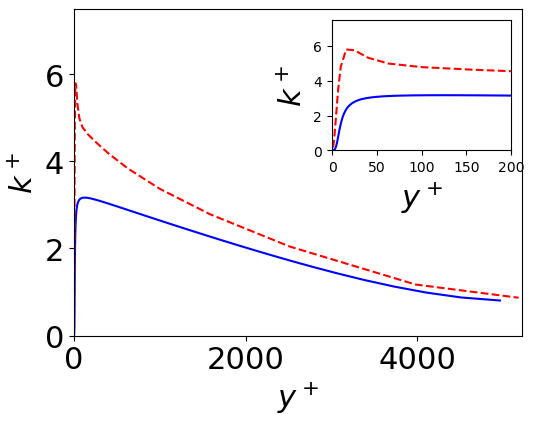}
\caption{Turbulent kinetic energy.}
\label{k-DNS}
\end{subfigure}
\begin{subfigure}[t]{0.5\textwidth}
\centering
\includegraphics[scale=0.28,clip=]{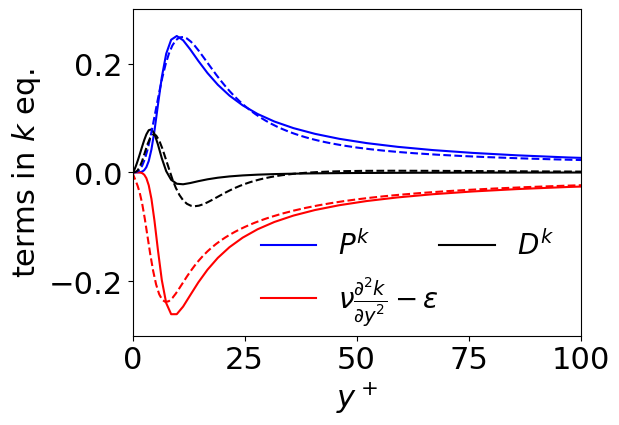}
\caption{Terms in the $k$ equation.  $D^k$ is turbulent diffusion; for the $k-\omega$
simulation $D^k  = \frac{\D}{\D y}\left(\nu_t \frac{\D k}{\D y}\right)$.}
\label{bal}
\end{subfigure}
\caption{Fully-developed channel flow.  $Re_\tau = 5\, 200$. 
Solid lines: $k-\omega$; dashed lines:DNS~\citep{moser:15}.}
\label{u-k-bal}
\end{figure}

In simple flows including a boundary layer (channel flow, flat-plate boundary layer flow, jet flow, etc)
the turbulent shear stress, $\uv$, must be correctly predicted. The turbulent shear stress is in these models 
linearly coupled to the turbulent viscosity, $\nu_t$, via
the Boussinesq assumption.
Hence, well-tuned eddy viscosity models correctly predict the mean flow,
the turbulent shear stress and the turbulent viscosity. However, the turbulent, kinetic energy, $k$,  is usually poorly predicted.
Figure~\ref{u-k-bal} presents a prediction using the Wilcox $k-\omega$ turbulence  model~\citep{wilcox:88} of fully
developed channel flow at $Re_\tau \equiv u_\tau \delta/\nu = 5\, 200$ where $u_\tau$, $\delta$ and $\nu$  
denote friction velocity, 
half-channel width and kinematic viscosity, respectively. It can be seen that the
$k-\omega$ model behaves as outlined above: the mean flow (and hence the turbulent shear stress and the turbulent viscosity) is well predicted
but the predicted turbulent kinetic energy is much too small. 

The object of the present paper is to improve the predicted, turbulent
kinetic energy. The predicted terms in the $k$ equation using the $k-\omega$ model are compared with DNS
 in Fig.~\ref{bal}. It can be seen that the production term, $P^k$,  and the sum of the viscous diffusion
and the dissipation term, $D^\nu - \e$,
 agree fairly well with DNS but that 
the agreement for the turbulent diffusion term, $D^k$,  is not so good. 
Regarding the sum of the viscous diffusion term and the dissipation
term, it may be noted that there is a slight spatial offset.

In order to improve the predicted $k$, we will use Physics Informed Neural Network, i.e. Physics Informed  NN -- 
usually  called PINN -- to improve the predicted diffusion term.  In \cite{yazdani:24} and \cite{luo:20} summarized
above, new constant values of turbulent coefficients were optimized. In the present study, one turbulent coefficient, $\sigma_k$,
is first turned into  a function of $y/\delta$, i.e. $\sigma_{k,PINN}\left(y/\delta\right)$. This coefficient appears
 in the diffusion term in the $k$ equation which in fully-developed channel flow reads
\begin{eqnarray}
\label{k-channel-flow}
\frac{d}{dy} \left[\left(\nu + \nu_{t,PINN}\right) \frac{dk}{dy}\right] +P^k - \e = 0
\end{eqnarray}
\begin{eqnarray}
\label{sigma_kPINN}
\sigma_{k,PINN} =\frac{\nu_t}{\nu_{t,PINN}}
\end{eqnarray}
and $\nu_{t} =k /\omega$.
The two last terms in Eq.~\ref{k-channel-flow} are the production and dissipation term, respectively.
The $\sigma_{k,PINN}$ function is obtained  as follows.
By taking the production, the dissipation terms as well as the turbulent kinetic energy from DNS channel flow 
data, Eq.~\ref{k-channel-flow} is turned into
an ordinary differential equation for the turbulent viscosity, $\nu_{t,PINN}$, which  is
solved using PINN. 

Recall that the turbulent viscosity is well predicted by the standard  Wilcox $k-\omega$ turbulence model. 
If we modify the turbulent kinetic energy, the turbulent viscosity will also be modified
and the predicted mean flow would then deteriorate. Hence, we must also modify the predicted $\omega$ so that the
new  $k-\omega$ model predicts the same turbulent viscosity, $\nu_t=k/\omega$,  as the Wilcox  $k-\omega$ model.
In order to not change the predicted
turbulent viscosity 
a new function, $C_{k,PINN}$, is added to the dissipation term in the $k$ equation and another, $C_{\omega 2,PINN}$, is added to
the destruction term in the $\omega$ equation. Finally,  three NN models are created for
$\sigma_{k,NN}$, $C_{k,NN}$ and $C_{\omega 2,NN}$ which are made
functions of two input parameters,  $\tau_{tot}/u_\tau^2$ and $\nu_t/(y u_\tau)$. The total stress, $\tau_{tot}$,  reads
\begin{equation}
\label{tau-tot}
\tau_{tot} = 2\left(\nu+\nu_t \right)  \left(\sbb_{ij}   \sbb_{ij}\right)^{1/2}, 
\quad \sbb_{ij} = \frac{1}{2} \left(\frac{\D \vb_i}{\D x_j} + \frac{\D \vb_j}{\D x_i}\right).
\end{equation}
The difference between $\sigma_{k,PINN}$ and $\sigma_{k,NN}$ is that the former is obtained
from  PINN and the latter is given by the NN model using $\sigma_{k,PINN}$ as the target.

The paper is organized as follows. In the two following sections, the two-dimensional solver (channel flows
and flat-plate boundary layer) and three-dimensional solver (periodic hill flow) are briefly
described. Then the PINN and NN models are presented. Next, the results are discussed and presented and in the final section we 
give some concluding remarks.

\section{Equations and the numerical method for the 2D RANS simulations}

The steady RANS equations used for the fully-developed channel flow and 
the flat-plate boundary layer read
\begin{eqnarray}
\label{contRANS}
\frac{\D \vb_i} {\D x_i}& =&  0\\
\label{RANS}
\frac{\D \left(\vb_i \vb_j\right)} {\D x_j} &=&  \delta_{1i}
- \frac{1}{\rho} \frac{\partial \bar{p}}{\D x_i}  + \frac{\D}{\D x_j}\left[(\nu + \nu_{t}) \frac{\D \vb_i}{\D x_j}\right] 
\end{eqnarray}
The first term on the right-hand side of Eq.~\ref{RANS}
 is the driving pressure gradient used in fully-developed channel flow; it is not present
in the flat-plate boundary layer.
The $k-\omega$ turbulence  model reads
\begin{equation}
\begin{split}
\label{kom}
\frac{\D\left( \vb_j k\right)}{\D x_j}  &=P^k+\frac{\D}{\D
x_j}\left[\left(\nu+\frac{\nu_t}{\sigma_{k,NN}}\right)\frac{\D k}{\D x_j}\right] -C_\mu C_{k,NN} k\omega\\
\frac{\D \left(\vb_j \omega\right)}{\D x_j}   &= C_{\omega_1}\frac{P^k}{\nu_t}
+ \frac{\D}{\D x_j}\left[\left(\nu+\frac{\nu_t}{\sigma_{\omega,NN}}\right)\frac{\D \omega}{\D x_j}\right] 
- C_{\omega 2, NN} \omega^{2} \\
P^k &= \nu_{t}\left(\frac{\D \vb_i}{\D x_j}+\frac{\D \vb_j}{\D x_i}\right) \frac{\D \vb_i}{\D x_j}, \quad 
\nu_t=\frac{k}{\omega}
\end{split}
\end{equation}
At the wall-adjacent cells, $\omega$ is not solved but is set as
\begin{eqnarray}
\label{omega_wallbc}
\omega_{w} =  \frac{6\nu}{C_{\omega 2, NN} y^2}
\end{eqnarray}
where $y$ is the wall distance between the wall-adjacent cell center and the wall. 
In the standard $k-\omega$ model $ C_{k,NN} = 1$, $\sigma_{k,NN} = \sigma_{k}$,  $\sigma_{\omega,NN}= \sigma_\omega$ 
and $ C_{\omega 2, NN} =  C_{\omega 2}$.
The coefficients have the values $C_{\omega 1}=5/9$,  $C_{\omega 2}=3/40$,
$\sigma_k=\sigma_\omega=2$ and $C_\mu=0.09$. 

It turns out that the turbulent Prandtl number, $\sigma_{k,NN}$, 
is near wall much smaller than $\sigma_k$. For that reason, also the turbulent Prandtl number
of the $\omega$ equation, $\sigma_\omega$, is modified as
\begin{equation}
\label{prandl_omeba}
\sigma_{\omega,NN} = \min\left(2\sigma_{k,NN},\sigma_\omega\right)
\end{equation}

The \pycalcr\ code is used~\citep{pycalc-rans-url} for solving the discretized equations. 
It is  an incompressible, finite volume code written in Python. It is fully vectorized
(i.e. no for loops).
The convective terms in the momentum equations are discretized using the MUSCL scheme~\citep{vanleer:79}
(a second-order bounded upwind scheme) and the hybrid $1^{st}$ order upwind/$2^{nd}$ order central scheme is used for 
$k$ and $\omega$ equations.
The numerical procedure is based on the pressure-correction method,
SIMPLEC,  and a collocated grid arrangement using Rhie-Chow interpolation~\citep{rhie:chow}.

\section{Equations and the numerical method for the hill-flow simulations}

The unsteady momentum equations for the periodic hill flow read
\begin{equation}
 \label{mom1} 
\frac{\D\vb_i}{\D t} + \frac{\D\left( \vb_j \vb_i\right)}{\D x_j} = \beta \delta_{1i}
- \frac{1}{\rho} \frac{\partial \bar{p}}{\D x_i}  + \frac{\D}{\D x_j}\left[(\nu + \nu_{t}) \frac{\D \vb_i}{\D x_j}\right] 
 \end{equation}
where the term $\beta \delta_{1i}$ is the driving pressure gradient in the streamwise direction.
The coefficient $\beta$ is given in Eq.~\ref{beta}.

The finite volume code \pycalc~\citep{pyCALC-LES} is used. It is written in Python and is
fully vectorized (i.e. no \texttt{for} loops).
 The solution procedure is based on fractional step. The second-order  MUSCL scheme~\citep{vanleer:79}
is used for the convective terms in the $\vb_i$ equations.
The $k$ and $\omega$ equations are given in Eq.~\ref{kom} (adding the unsteady term $\D /\D t$) using the same
discretization.

The only reason why \pycalcr\ is not used for this flow is that the author did not succeed in adjusting the $\beta$
coefficient, in a reasonable manner, see Eq.~\ref{beta}. The discretization in space in \pycalc\ is identical to that
in \pycalcr. The main difference is how the pressure-velocity coupling  is treated; furthermore  the latter code was 
developed for unsteady,  three-dimensional flow  whereas the former is used for steady,  two-dimensional flow.

Above, we have used tensor notation for the velocity vector ($\bar{v}_1, \bar{v}_2, \bar{v}_3$) and for the
space vector ($x_1, x_2, x_3$). Below, we will replace them 
with  $\bar{u}, \bar{v}, \bar{w}$ and  $x, y, z$.

\section{Development of the new $k-\omega$ model using PINN and NN}

The implementation of PINN is briefly explained in Appendix~\ref{PINN}.
Now, let's involve our differential equation, the $k$  equation (see Eq.~\ref{kom}). The equation for turbulent
kinetic energy in fully-developed channel flow is given in Eq.~\ref{k-channel-flow}
and it is re-written as
\begin{eqnarray}
\label{ODE}
\left (\nu + \nu_{t,PINN}\right)  \frac{d^2k}{dy^2} +  \frac{dk}{dy}  \frac{d\nu_{t,PINN}}{dy} +  P^k - \e = Q
\end{eqnarray}
where we have added a right-hand side, $Q$ (which is an error term that should be reduced to zero, see below). 
We want to find a new turbulent viscosity, $\nu_{t,PINN}$, that gives a turbulent diffusion 
that agrees with the DNS turbulent diffusion
term in Fig.~\ref{bal}. Hence, 
$\nu_{t,PINN}$ is the unknown variable in Eq.~\ref{ODE} 
and $k$, $P^k$ and $\e$ are taken from DNS. Equation~\ref{ODE} is solved in a half-width channel at $Re_\tau = 5\, 200$.
The boundary condition at the wall ($y=0$) is $\nu_{t,PINN} = 0$ and at the center ($y=\delta$) it is $\nu_{t,PINN} = \nu_{t,DNS}$
where $ \nu_{t,DNS} = \left |\frac{\overline{u'v'}}{\D \vb/\D y}\right|_{DNS}$.

\begin{figure}
\centering
\begin{subfigure}[t]{0.5\textwidth}
\centering
\includegraphics[scale=0.28,clip=]{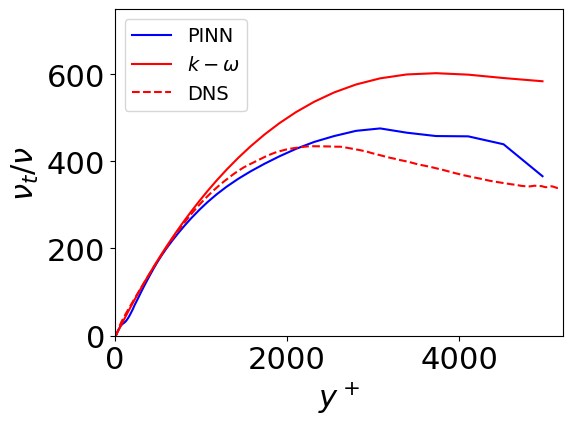}
\caption{Turbulent viscosity. Full view}
\label{pinn-vist}
\end{subfigure}%
\begin{subfigure}[t]{0.5\textwidth}
\centering
\includegraphics[scale=0.28,clip=]{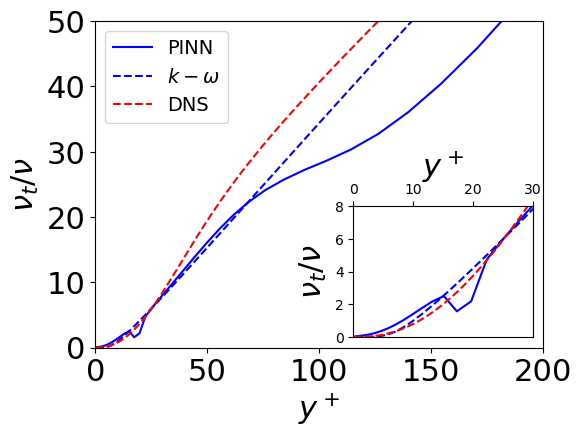}
\caption{Turbulent viscosity. Zoomed view}
\label{pinn-vist-zoom}
\end{subfigure}
\begin{subfigure}[t]{0.5\textwidth}
\centering
\includegraphics[scale=0.28,clip=]{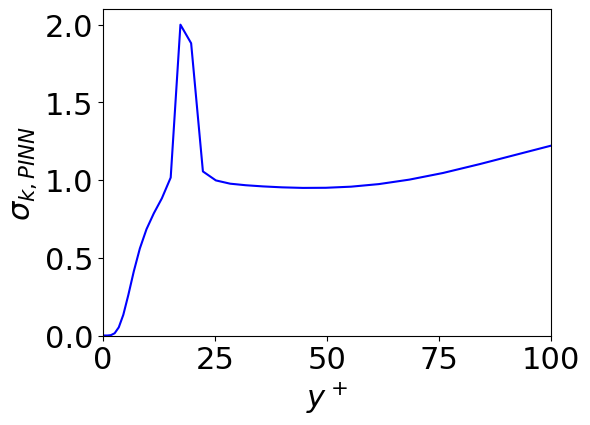}
\caption{Prandtl number.}
\label{pinn-prandtl}
\end{subfigure}%
\begin{subfigure}[t]{0.5\textwidth}
\centering
\includegraphics[scale=0.28,clip=]{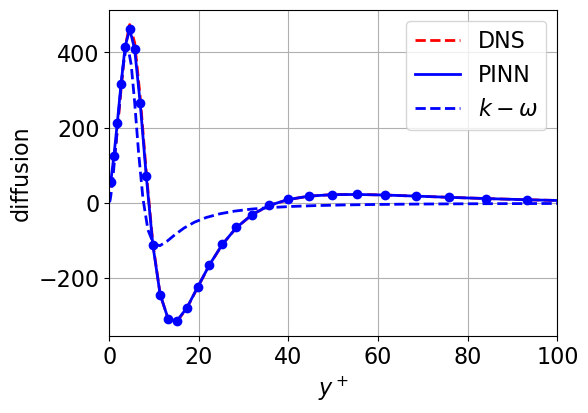}
\caption{Turbulent diffusion.}
\label{pinn-diff}
\end{subfigure}
\caption{Prediction by PINN compared to DNS and the Wilcox $k-\omega$ model. $Re_\tau = 5\, 200$.}
\end{figure}

\begin{figure}
\centering
\includegraphics[scale=0.28,clip=]{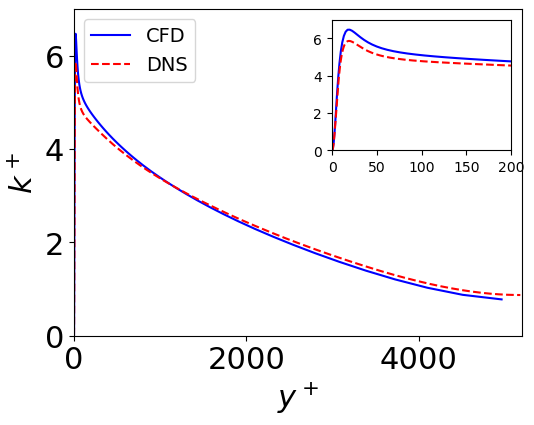}
\caption{CFD prediction of Eq.~\ref{ODE-k} using $\sigma_{k,PINN}$.}
\label{ODE-k-fig}
\end{figure}

\begin{figure}
\centering
\includegraphics[scale=0.28,clip=]{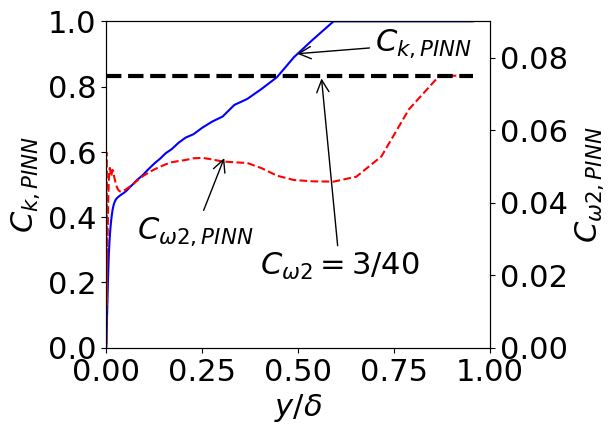}
\caption{$C_{k,PINN}$ and $C_{\omega 2,PINN}$.}
\label{c_k-c_omega_2}
\end{figure}

\begin{figure}
\centering
\begin{subfigure}[t]{0.5\textwidth}
\centering
\includegraphics[scale=0.28,clip=]{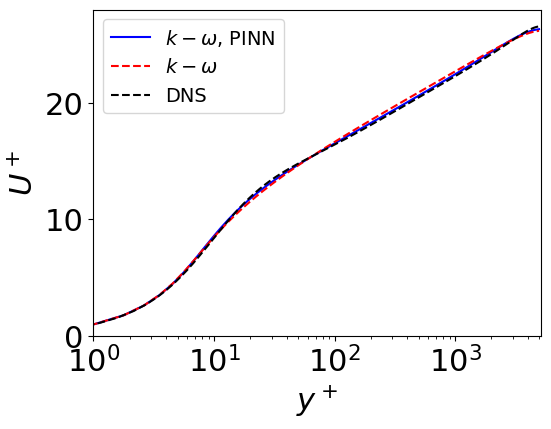}
\caption{Velocity.}
\label{vel-5200-PINN}
\end{subfigure}%
\begin{subfigure}[t]{0.5\textwidth}
\centering
\includegraphics[scale=0.28,clip=]{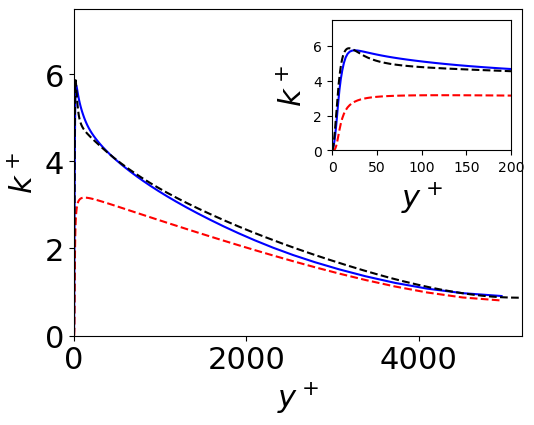}
\caption{Turbulent  kinetic energy.}
\label{k-5200-PINN}
\end{subfigure}
\begin{subfigure}[t]{0.5\textwidth}
\centering
\includegraphics[scale=0.28,clip=]{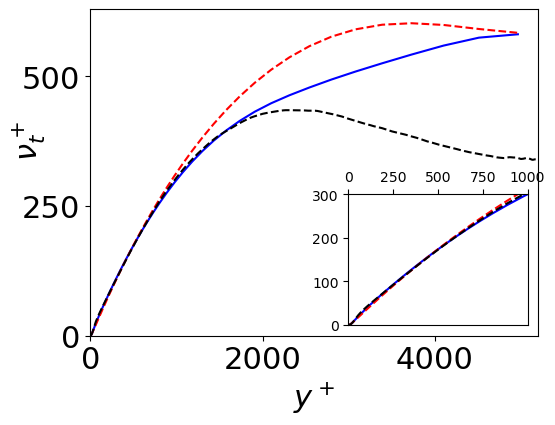}
\caption{Turbulent viscosity.}
\label{vist-5200-PINN}
\end{subfigure}%
\caption{Fully-developed channel flow comparing $k-\omega$ PINN and standard $k-\omega$ model. $\sigma_{k,PINN}$
(Eqs.~\ref{ODE} and \ref{sigma_kPINN}),
$C_{k,PINN}$ (Eq.~\ref{C_k}) and $C_{\omega 2, PINN}$ (Eq.~\ref{C_omega}) are used in Eq.~\ref{kom}.  $Re_\tau = 5\,200$.}
\label{channel-5200-PINN}
\end{figure}

The turbulent viscosity, $\nu_{t,PINN}$  in Eq.~\ref{ODE}, will be predicted by PINN while minimizing 
the error $Q^2$.
The loss function, {\tt loss\_fn}, in Listing~\ref{NN_0_list-code}  in the Appendix
 is replaced with Eq.~\ref{ODE} and the Python code is given in 
Listing~\ref{PINN_0}.
\texttt{nut} and \texttt{k} in Listing~\ref{PINN_0} are $\nu_{t,PINN}$ and $k$ in Eq.~\ref{ODE}, respectively. 
\texttt{k\_y}, for example, 
is $dk/dy$. Note that \texttt{k\_y} and  \texttt{k\_yy} are known (taken from DNS). There  are two losses in  Listing~\ref{PINN_0},
one, \texttt{ODE\_loss},  which relates to the ordinary differential equation (ODE)
 and one,  \texttt{BC\_loss},  which relates to the boundary conditions (BC).
In this
way the PINN is forced to satisfy both the ODE and the boundary conditions. It should be mentioned that we had serious problems
in making the PINN converge. The first problem was that the PINN is sensitive
to the initialization of the weights and biases. Two subsequent runs gave very different convergence histories.
The remedy was to save the initial  weights and biases of a successful
run and then  load and use them in subsequent runs.  The second remedy was to increase the number of neurons from $10$ to $30$,  
switch from the \texttt{tanh} actuator to the \texttt{sigmoid} actuator and increase the number of hidden layers from $3$ to $5$, 
see \texttt{class MyNet} in Listing~\ref{PINN_0}. The Adam  optimizer was used with an initial learning rate
of $0.01$ which was reduced by $0.7$ (\texttt{gamma=0.7}) using 
the \texttt{MultiStepLR} scheduler at
\texttt{milestones} equal to \texttt{3e4, 6e4, 8e4, 1e5, 1.3e5, 1.5e5, 1.8e5}. This scheduler was found to be more
suitable than the \texttt{ReduceLROnPlateau} since it was difficult to find a good \texttt{patience} parameter (i.e.
how long a plateau should be defined before the learning rate should be reduced). The loss, defined as $\sum (Y-y_{pred})^2$
is during $200\, 000$ epochs reduced from $1.4\cdot 10^{13}$ to $4$.
 Recall that all Python scripts can be downloaded~\citep{davidson-pinn-nn:25}.

In the  NN models, contrary to the PINN solver, the \texttt{relu} actuator function was found to be superior
to the \texttt{sigmoid} actuator function.

Figures~\ref{pinn-vist} - \ref{pinn-prandtl}  present the predicted turbulent viscosity in the $k$ equation, $\nu_{t,PINN}$,  and 
the turbulent Prandtl number, $\sigma_{k,PINN}$,  
where $\sigma_{k,PINN} = \nu_t/\nu_{t,PINN}$.  It is seen that
the PINN turbulent viscosity for $y^+ \lesssim 15$ is slightly larger than that predicted by the $k-\omega$ model.
But it should be noted that the object of PINN is not that $\nu_{t,PINN}$ should agree 
with that of  the $k-\omega$ model
or DNS but
to predict a turbulent diffusion that agrees with DNS (see Fig.~\ref{pinn-diff}). 
The DNS turbulence diffusion includes the contribution from both the triple correlation and the pressure-velocity
correlation; the latter term was (by mistake) omitted in \cite[Fig.~3c]{davidson_etmm15}.
The DNS and PINN lines in  Fig.~\ref{pinn-diff} are on top of each other.
In the inset in 
Fig.~\ref{pinn-vist-zoom} a small non-physical oscillation can be seen. It occurs at the point where $d k_{DNS}/dy$ changes sign.

It should be noted that $\nu_{t,PINN}$ in Fig.~\ref{pinn-vist-zoom} is for  $y^+ \gtrsim 30$
very different to that shown in~\cite[Fig.~3a]{davidson_etmm15} .
 The reason is that Eq.~\ref{ODE} was poorly converged for $y^+ \gtrsim 30$ in~\cite{davidson_etmm15}.
 However, the predicted turbulent
diffusion in~\cite[Fig.~3c]{davidson_etmm15} is very similar to that in Fig.~\ref{pinn-diff}.

Let's verify that a CFD solver does predict a correct
turbulent kinetic energy using the PINN turbulent Prandtl number (see Fig.~\ref{pinn-prandtl}). 
In  fully-developed channel flow, the $k$ equation (see Eq.~\ref{kom}) reads
\begin{equation}
\label{ODE-k}
\frac{d}{dy}\left(\left(\nu + \frac{\nu_t}{\sigma_{k,PINN}}\right) \frac{dk}{dy}\right)  +P^k - \e = 0
\end{equation}
where $P^k$ and $\e$ are again taken from DNS. The turbulent viscosity, $\nu_{t}$, 
is computed as  $\nu_t = k/\omega_{DNS}$. $\omega_{DNS}$ is taken from Eq.~\ref{nut} since
the turbulence viscosity, $\nu_t$,
must be the same as that in the $k-\omega$ model (and DNS), see Eq.~\ref{nut}.

We solve Eq.~\ref{ODE-k} using the CFD solver \pycalcr. Figure~\ref{ODE-k-fig} presents the predicted
$k$ profiles. We find that the agreement is excellent.

Until now  we have modified the turbulent Prandtl number in the $k$ equation 
so that we -- with exact source terms taken from DNS, $P^k$ and
$\e$ -- predict a correct near-wall behaviour of $k$. In the next step, we will solve both the $k$ and $\omega$ equations
computing the source terms as in Eq.~\ref{kom}. 
Recall that $P^k$ in the Wilcox $k-\omega$ model is correct since the model does predict the velocity
profile correctly, see Fig.~\ref{u-k-bal},  and hence it also predicts
 the turbulent viscosity correctly. Recall that the total shear stress is predicted as 
\begin{equation}
\label{y-1}
y/\delta-1
\end{equation}
by any turbulence models in a finite volume method since this is the exact solution when integrating the momentum 
equation~\cite[Section 6.2]{davidson:MoF-url}.
 The $k$ predicted by DNS near the wall is much larger than that predicted 
by the Wilcox $k-\omega$ model, see Fig.~\ref{k-DNS}. This means that $\omega$ must be modified in order to give the same turbulent viscosity
as the Wilcox $k-\omega$ model, i.e. 
\begin{equation}
\label{nut}
\nu_{t,k-\omega} = \frac{k_{k-\omega}}{\omega_{k-\omega}} = \nu_{t,DNS} =  \frac{k_{DNS}}{\omega_{DNS}}
\end{equation}
which gives
\begin{equation}
\label{omega_DNS}
\omega_{DNS} =  k_{DNS}/\nu_{t,k-\omega}.
\end{equation}

Now we have an $\omega_{DNS}$ that gives the same turbulent viscosity as DNS. 
Next, the dissipation term $C_\mu k \omega$ 
in the $k$ equation in Eq.~\ref{kom} must be modified so that it agrees with $\e_{DNS} = C_\mu k_{DNS}\omega_{DNS}$. 
This is achieved by
multiplying the dissipation term by a damping function, $C_{k,PINN} = C_{k,PINN}\left(y/\delta\right)$, 
 so that the $k$ equation  in Eq.~\ref{kom} is satisfied with $k=k_{DNS}$ and $\omega = \omega_{DNS}$, 
i.e. (see Eq.~\ref{ODE-k})
\begin{equation}
\label{C_k}
C_{k,PINN} = \frac{D^k_{DNS}   +P^k_{DNS}}{ C_\mu k_{DNS} \omega_{DNS}} 
\end{equation}
Note that the viscous diffusion has been omitted in order to prevent a large gradient of $C_{k,PINN}$ close to the wall.

Finally, we must make sure that the $\omega$ equation in Eq.~\ref{kom} 
predicts $\omega = \omega_{DNS}$. This is achieved by making $C_{\omega 2,PINN}$ = $C_{\omega 2,PINN}\left(y/\delta\right)$.
From the $\omega$ equation in Eq.~\ref{kom} we get 
\begin{equation}
\label{C_omega}
C_{\omega 2,PINN} = \frac{\frac{d}{dy}\left(\frac{\nu_t}{\sigma_{\omega}} \frac{d\omega_{DNS}}{dy}\right)  + 
C_{\omega 1} \frac{P^k_{DNS}}{\nu_{t,DNS}}}{\omega^2_{DNS}} 
\end{equation}
Again, the viscous diffusion has been omitted in 
order to prevent a large gradient of $C_{\omega 2,PINN}$ close to the wall.

\begin{figure}
\centering
\begin{subfigure}[t]{0.5\textwidth}
\centering
\includegraphics[scale=0.28,clip=]{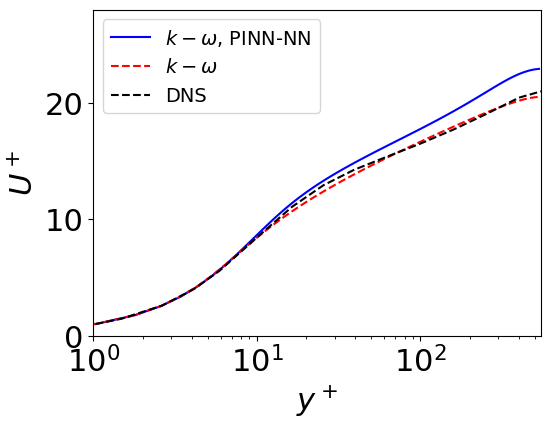}
\caption{Velocity.}
\label{vel-550}
\end{subfigure}%
\begin{subfigure}[t]{0.5\textwidth}
\centering
\includegraphics[scale=0.28,clip=]{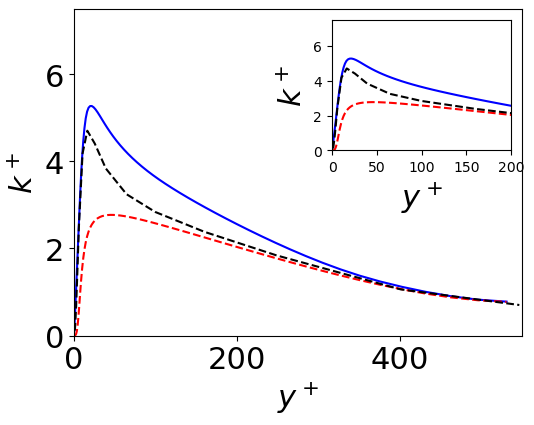}
\caption{Turbulent  kinetic energy.}
\label{k-550}
\end{subfigure}
\begin{subfigure}[t]{0.5\textwidth}
\centering
\includegraphics[scale=0.28,clip=]{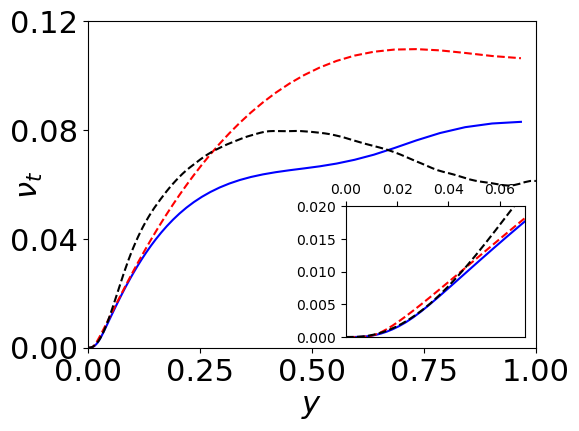}
\caption{Turbulent viscosity.}
\label{vist-550}
\end{subfigure}%
\begin{subfigure}[t]{0.5\textwidth}
\centering
\includegraphics[scale=0.28,clip=]{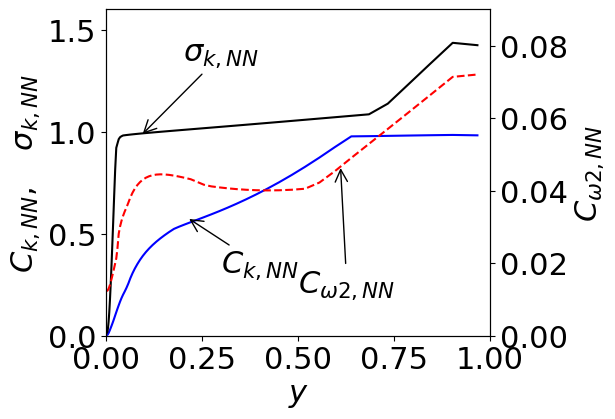}
\caption{$\sigma_{k,NN}$, $C_{k,NN}$, $C_{\omega 2,NN}$ predicted by the NN model.}
\label{prand_k-550}
\end{subfigure}
\begin{subfigure}[t]{\textwidth}
\centering
\includegraphics[scale=0.28,clip=]{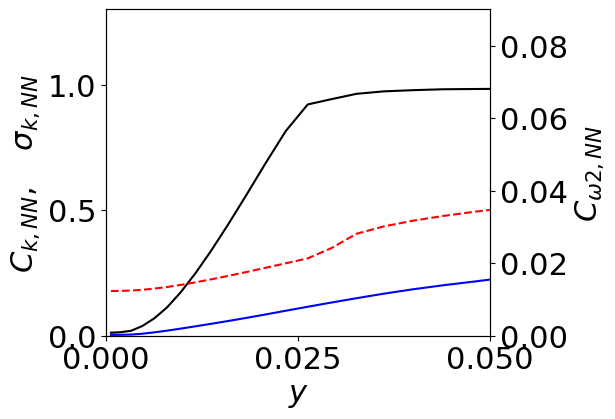}
\caption{$\sigma_{k,NN}$, $C_{k,NN}$, $C_{\omega 2,NN}$ predicted by the NN model. Zoomed-in view.}
\label{prand_k-550-zoom}
\end{subfigure}
\caption{Fully-developed channel flow. $Re_\tau = 550$.}
\label{channel-550}
\end{figure}

\begin{figure}
\centering
\begin{subfigure}[t]{0.5\textwidth}
\centering
\includegraphics[scale=0.28,clip=]{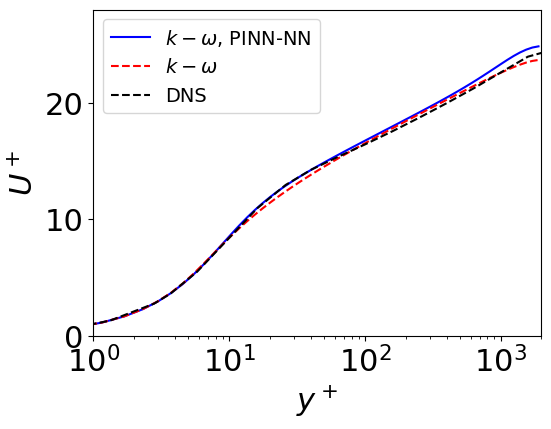}
\caption{Velocity.}
\label{vel-2000}
\end{subfigure}%
\begin{subfigure}[t]{0.5\textwidth}
\centering
\includegraphics[scale=0.28,clip=]{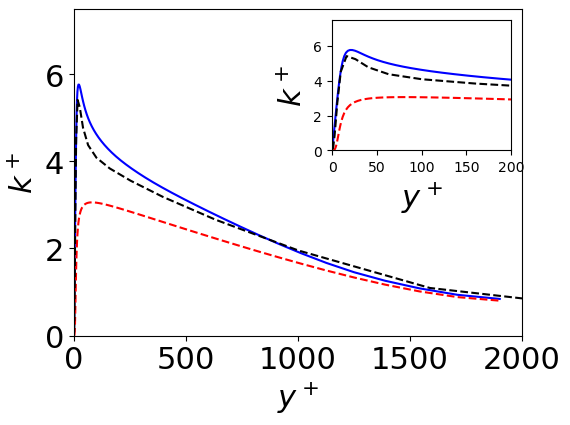}
\caption{Turbulent  kinetic energy.}
\label{k-2000}
\end{subfigure}
\begin{subfigure}[t]{0.5\textwidth}
\centering
\includegraphics[scale=0.28,clip=]{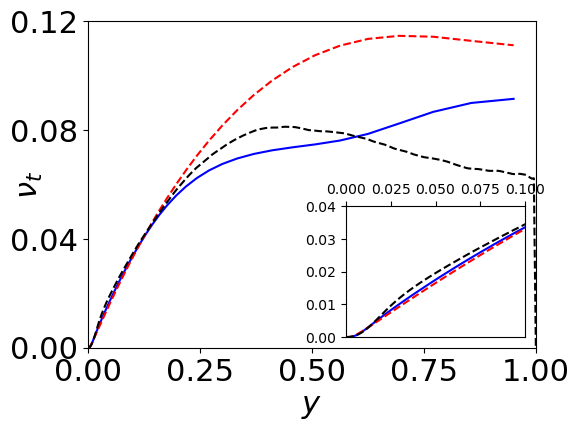}
\caption{Turbulent viscosity.}
\label{vist-2000}
\end{subfigure}%
\begin{subfigure}[t]{0.5\textwidth}
\centering
\includegraphics[scale=0.28,clip=]{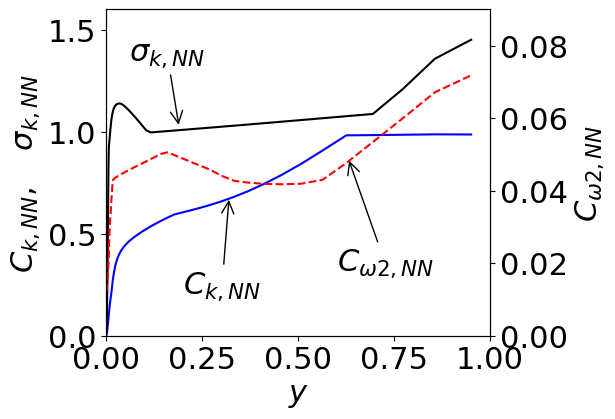}
\caption{$\sigma_{k,NN}$, $C_{k,NN}$, $C_{\omega 2,NN}$ predicted by the NN model.}
\label{prand_k-2000}
\end{subfigure}
\begin{subfigure}[t]{\textwidth}
\centering
\includegraphics[scale=0.28,clip=]{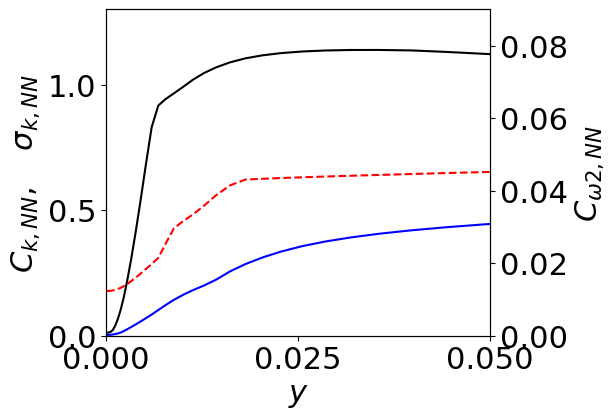}
\caption{$\sigma_{k,NN}$, $C_{k,NN}$, $C_{\omega 2,NN}$ predicted by the NN model. Zoomed-in view.}
\label{prand_k-2000-zoom}
\end{subfigure}
\caption{Fully-developed channel flow. $Re_\tau = 2\, 000$.}
\label{channel-2000}
\end{figure}

\begin{figure}
\centering
\begin{subfigure}[t]{0.5\textwidth}
\centering
\includegraphics[scale=0.28,clip=]{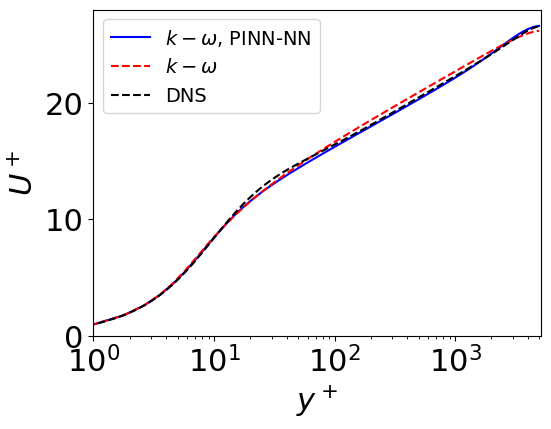}
\caption{Velocity.}
\label{vel-5200}
\end{subfigure}%
\begin{subfigure}[t]{0.5\textwidth}
\centering
\includegraphics[scale=0.28,clip=]{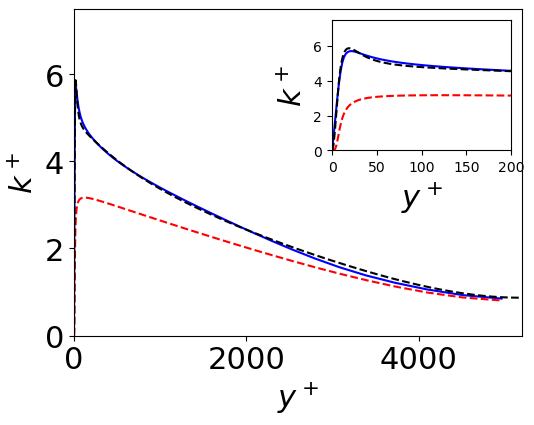}
\caption{Turbulent  kinetic energy.}
\label{k-5200}
\end{subfigure}
\begin{subfigure}[t]{0.5\textwidth}
\centering
\includegraphics[scale=0.28,clip=]{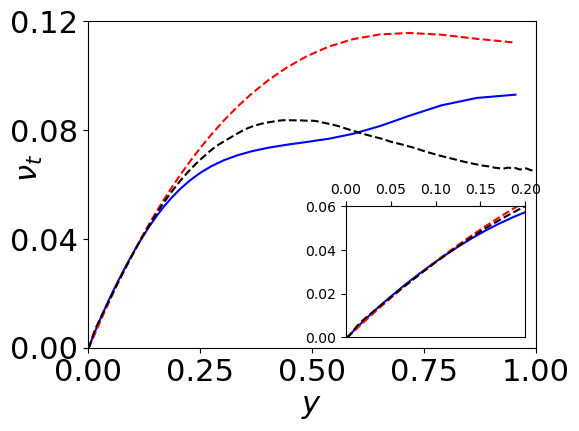}
\caption{Turbulent viscosity.}
\label{vist-5200}
\end{subfigure}%
\begin{subfigure}[t]{0.5\textwidth}
\centering
\includegraphics[scale=0.28,clip=]{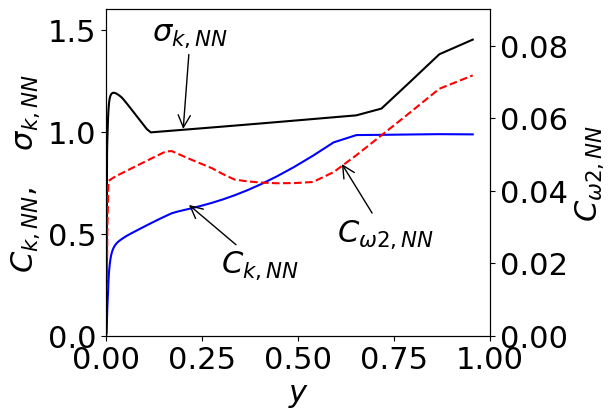}
\caption{$\sigma_{k,NN}$, $C_{k,NN}$, $C_{\omega 2,NN}$ predicted by the NN model.}
\label{prand_k-5200}
\end{subfigure}
\begin{subfigure}[t]{\textwidth}
\centering
\includegraphics[scale=0.28,clip=]{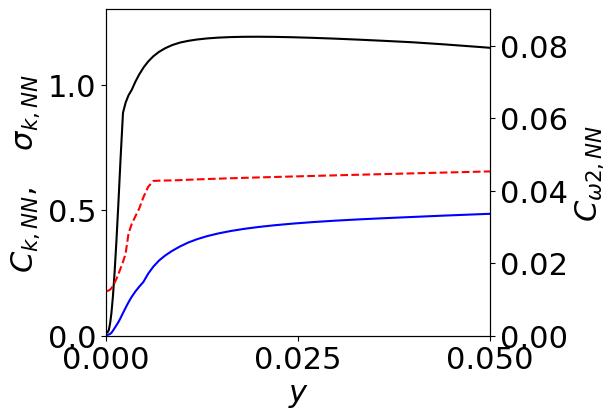}
\caption{$\sigma_{k,NN}$, $C_{k,NN}$, $C_{\omega 2,NN}$ predicted by the NN model. Zoomed-in view.}
\label{prand_k-5200-zoom}
\end{subfigure}
\caption{Fully-developed channel flow. $Re_\tau = 5\, 200$.}
\label{channel-5200}
\end{figure}

\begin{figure}
\centering
\begin{subfigure}[t]{0.5\textwidth}
\centering
\includegraphics[scale=0.28,clip=]{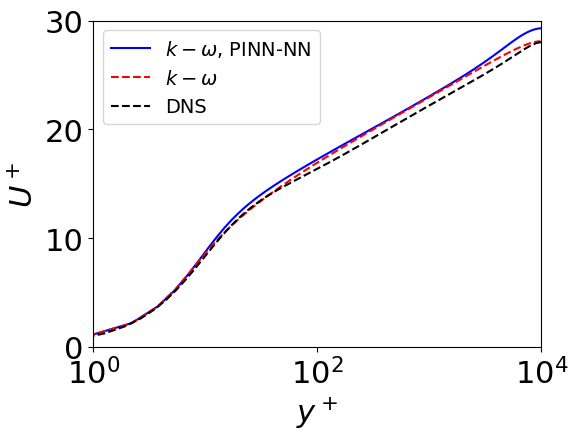}
\caption{Velocity.}
\label{vel-10000}
\end{subfigure}%
\begin{subfigure}[t]{0.5\textwidth}
\centering
\includegraphics[scale=0.28,clip=]{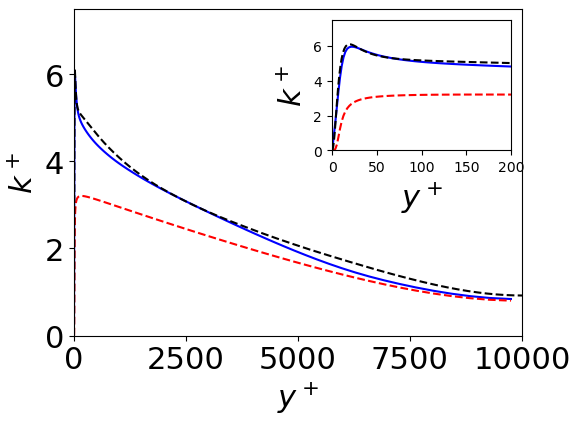}
\caption{Turbulent  kinetic energy.}
\label{k-10000}
\end{subfigure}
\begin{subfigure}[t]{0.5\textwidth}
\centering
\includegraphics[scale=0.28,clip=]{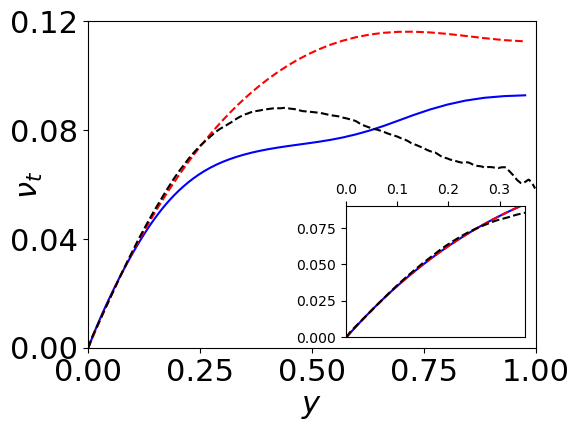}
\caption{Turbulent viscosity.}
\label{vist-10000}
\end{subfigure}%
\begin{subfigure}[t]{0.5\textwidth}
\centering
\includegraphics[scale=0.28,clip=]{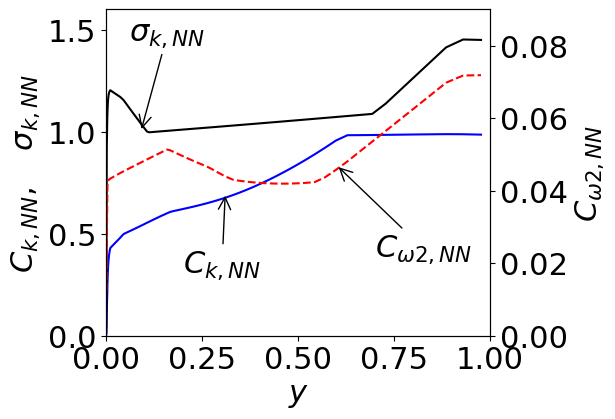}
\caption{$\sigma_{k,NN}$, $C_{k,NN}$, $C_{\omega 2,NN}$ predicted by the NN model.}
\label{prand_k-10000}
\end{subfigure}
\begin{subfigure}[t]{\textwidth}
\centering
\includegraphics[scale=0.28,clip=]{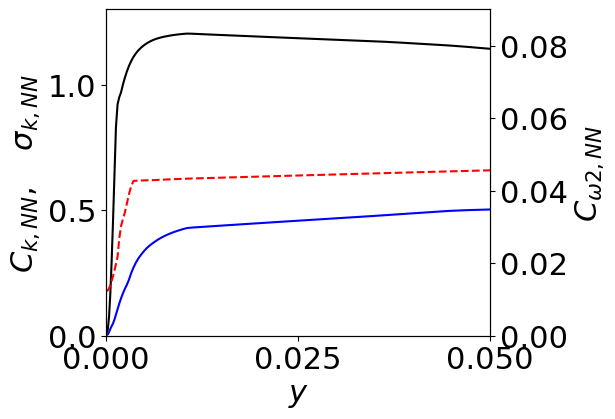}
\caption{$\sigma_{k,NN}$, $C_{k,NN}$, $C_{\omega 2,NN}$ predicted by the NN model. Zoomed-in view.}
\label{prand_k-10000-zoom}
\end{subfigure}
\caption{Fully-developed channel flow. $Re_\tau = 10\, 000$.}
\label{channel-10000}
\end{figure}

\begin{figure}
\centering
\includegraphics[scale=0.28,clip=]{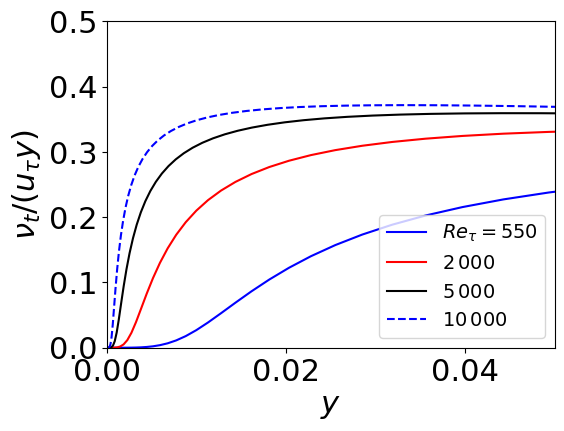}
\caption{Fully-developed channel flow. Near-wall behavior of the input paramenter $\nu_t/(u_\tau y)$.}
\label{nut-over-y}
\end{figure}

\begin{figure}
\centering
\includegraphics[scale=0.28,clip=]{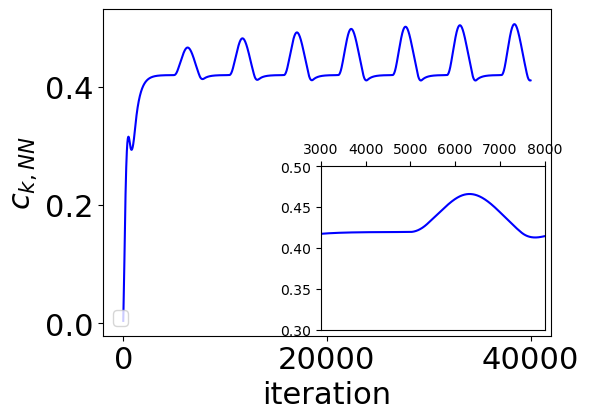}
\caption{Fully-developed channel flow. $Re_\tau = 10\, 000$.
$C_{k,NN}$ vs. iteration at $y^+ = 100$. $C_{k,NN}$ has been filtered over $500$ iterations }
\label{c_k-iter-10000}
\end{figure}

\begin{table}
\begin{center}
\begin{tabular}{|c|c|c|c|c|c|}\hline
$Re_\tau$ & $N_y$ & $s_y$ & $ y^+$ \\ \hline
$ 550$  & $60$ & $1.07$ & $ 0.3$ \\ \hline
$2000$ & $60$ & $1.11$ & $0.2$ \\ \hline
$5\, 200$ & $70$ & $1.1$ & $0.3$ \\ \hline
$10\, 000$   &  $150$  & $1.05$   & $0.2$  \\ \hline
\end{tabular}
\end{center}
\caption{Channel flow. Reynolds number ($Re_\tau$), number of cells in the wall-normal direction ($N_y$),
grid stretching ($s_y$) and location of wall-adjacent cell center ($ y^+$).}
\label{tabell-channel}
\end{table}

\begin{figure}
\centering
\begin{subfigure}[t]{0.5\textwidth}
\centering
\includegraphics[scale=0.28,clip=]{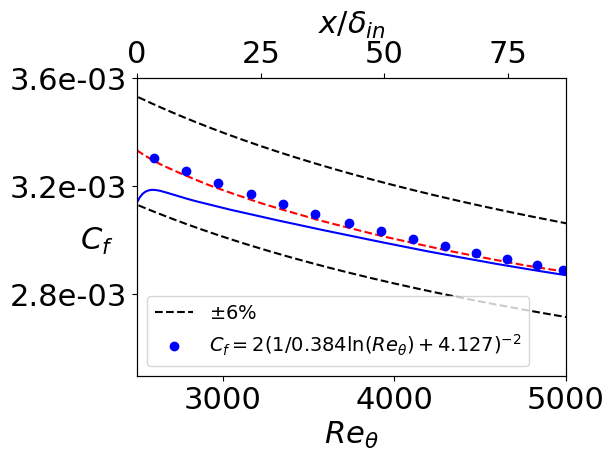}
\caption{Skin friction.}
\label{el-bound}
\end{subfigure}%
\begin{subfigure}[t]{0.5\textwidth}
\centering
\includegraphics[scale=0.28,clip=]{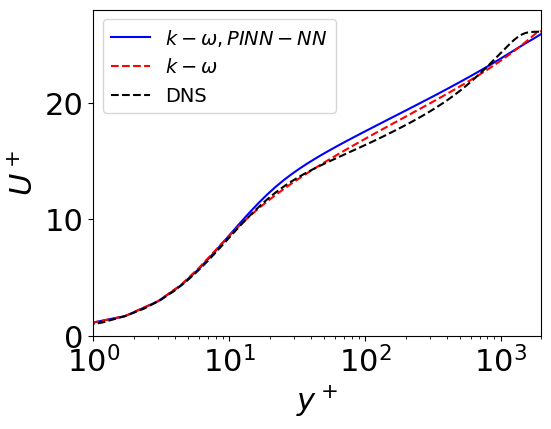}
\caption{Velocity.}
\label{vel-bound}
\end{subfigure}
\begin{subfigure}[t]{0.5\textwidth}
\centering
\includegraphics[scale=0.28,clip=]{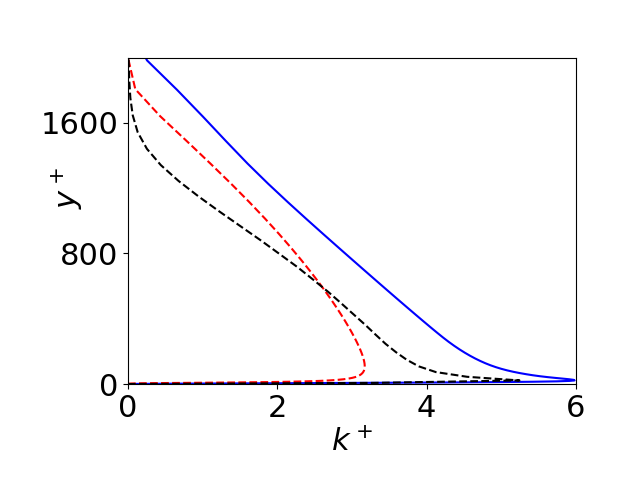}
\caption{Turbulent  kinetic energy.}
\label{k-bound}
\end{subfigure}%
\begin{subfigure}[t]{0.5\textwidth}
\centering
\includegraphics[scale=0.28,clip=]{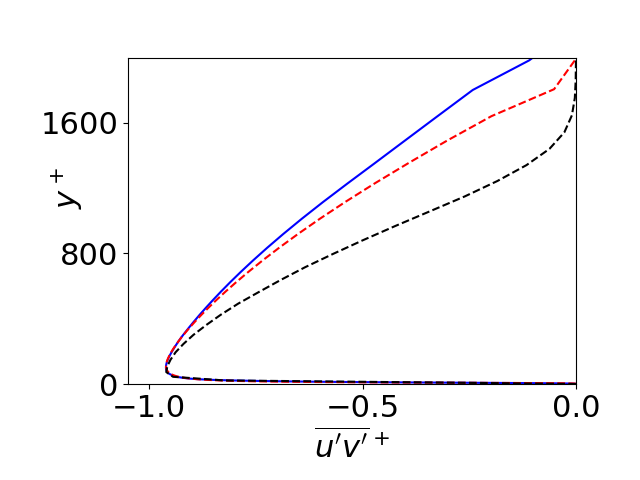}
\caption{Turbulent shear stress.}
\label{vist-bound}
\end{subfigure}
\begin{subfigure}[t]{0.5\textwidth}
\centering
\includegraphics[scale=0.28,clip=]{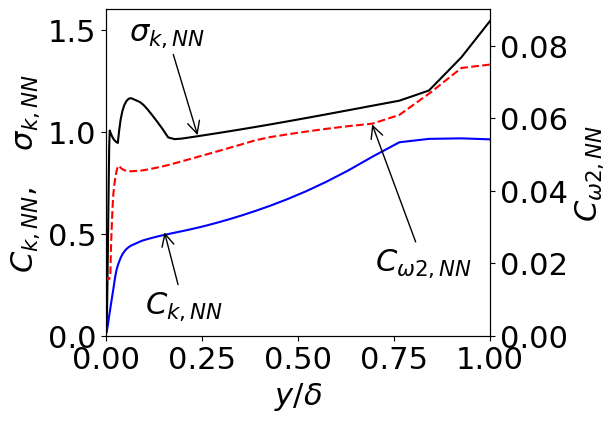}
\caption{$\sigma_{k,NN}$, $C_{k,NN}$, $C_{\omega 2,NN}$ predicted by the NN model.}
\label{prand_k-bound}
\end{subfigure}
\caption{Flat-plate boundary layer. Profiles at $Re_\theta = 4\, 500$. DNS~\citep{sillero:13}}
\label{bound}
\end{figure}

Figure~\ref{c_k-c_omega_2} shows the two damping functions computed with 
 Eqs.~\ref{C_k} and \ref{C_omega}. The thick, black, dashed line
is the standard value of $C_{\omega, 2}=3/40$, see below Eq.~\ref{kom}. 
The damping function, $C_{k,PINN}$, is in the outer region much affected by the dominator in Eq.~\ref{C_k} which  essentially is
 $\left(k_{DNS}/k_{k-\omega}\right)^2$: first,
$k_{DNS}$ is larger than $k_{k-\omega}$ (see Fig.~\ref{ODE-k-fig}), and, second, $\omega_{DNS}$ is larger than $\omega_{k-\omega}$, 
see Eq.~\ref{nut}. This explains why $C_{k,PINN}$ is much smaller than one  for $y/\delta \lesssim 0.5$.

Finally, we insert our  expression of $\sigma_{k,PINN}$ (obtained by PINN, Eq.~\ref{ODE} and Eq.~\ref{sigma_kPINN}), 
$C_{k,PINN}$ (Eq.~\ref{C_k})  and $C_{\omega 2,PINN}$  (Eq.~\ref{C_omega}) into Eq.~\ref{kom} and solve the
full equation system ($\vb_1$, $k$ and $\omega$) using \pycalcr. The predictions are compared with DNS in Fig.~\ref{channel-5200-PINN}
and we find that the agreement is very good.

\subsection{Compute $\sigma_{k,NN}$, $C_{k,NN}$ and  $C_{\omega 2,NN}$ using Neural Network (NN)}

\label{sect:NN}

It was shown~\cite{davidson_etmm15} that the $k-\omega$ model, using  $\sigma_{k,PINN}$, $C_{k,PINN}$ and  $C_{\omega 2,PINN}$, 
accurately  predicts fully-developed channel flow 
at $Re_\tau = 2\, 000$,  $Re_\tau = 5\, 200$, 
  $Re_\tau = 10\, 000$ as well as a flat-plate boundary layer.
But the drawback is that this model is not applicable to re-circulating flow since the three coefficients are expressed in $y/\delta$.
In this section, we will use NN models  to predict  $\sigma_{k,NN}$, $C_{k,NN}$ and $C_{\omega 2,NN}$. 
Many different input parameters to 
the NN models were investigated such as $P_k/\e$, $P^+_k$, $\e^+$, $\nu_t/(y u_\tau)$, etc,
It is important that they are properly non-dimensionalized so that they can be used in other flows
as well at other Reynolds numbers.
In the end, a good combination was found:  the input parameters to the NN models 
for     $\sigma_{k,NN}$, $C_{k,NN}$ and $C_{\omega 2,NN}$  were taken as 
\begin{equation}
\frac{\tau_{tot}}{u_\tau^2}  \quad \textrm{and} \quad \frac{\nu_t}{y u_\tau}
\end{equation}
where $y$ is the distance to the nearest wall and $\tau_{tot}$ is given in Eq.~\ref{tau-tot}.
Using these input parameters the NN models will be applicable to complex, recirculating flow.
In complex geometries with multiple bodies, it may be costly to find the wall distance to the closest wall.
An efficient to find the wall distance is to solve a Poisson equation~\citep{tucker:98}.
The input parameters, $\tau_{tot}/u_\tau^2$ and $\nu_t/(y u_\tau)$, 
are in the training process taken from the $k-\omega$-PINN predictions in Fig.~\ref{channel-5200-PINN}.
The targets/outputs are $\sigma_{k,PINN}$ (Eq.~\ref{sigma_kPINN}), $C_{k,PINN}$ (Eq.~\ref{C_k})
  and $C_{\omega 2,PINN}$ (Eq.~\ref{C_omega}). It should be mentioned that the non-physical oscillation
in Fig.~\ref{pinn-vist-zoom}  was smoothed when  $\sigma_{k,PINN}$ was used as target in the NN model for  $\sigma_{k,NN}$
below.

Minimum and maximum of the input parameters, $\tau_{tot}/u_\tau^2$ and $\nu_t/(y u_\tau)$, 
as well as the output parameters, $\sigma_{k,NN}$, $C_{k,NN}$,  $C_{\omega 2,NN}$,
are in the training process stored on disk. They are then used in the NN models
in the CFD code to set limits on the CFD input parameters, see lines 42-49 in  Listing~\ref{PINN_NN} in the Appendix,
and the output parameters, see lines 65-67.
The difference between $\sigma_{k,PINN}$ and $\sigma_{k,NN}$ is that the former is obtained from Eq.~\ref{sigma_kPINN}
where $\nu_{t,PINN}$ is given by solving an ODE (Eq.~\ref{ODE-k}) using PINN; the latter is predicted by the NN model
using  $\sigma_{k,PINN}$ as the target. The differences between  $C_{k,PINN}$/$C_{k,NN}$ and
$C_{\omega 2,PINN}$/$C_{\omega 2,NN}$ are essentially the same: $C_{k,PINN}$  and $C_{\omega 2,PINN}$ are  given by
Eqs.~\ref{C_k} and \ref{C_omega} whereas $C_{k,NN}$  and $C_{\omega 2,NN}$ 
are predicted by the NN models using $C_{k,PINN}$  and $C_{\omega 2,PINN}$ as targets.

All Python PINN and NN  
 scripts and the Python CFD codes can be downloaded~\citep{davidson-pinn-nn:25}.

\begin{figure}
\centering
\begin{tikzpicture}[xscale=0.6, yscale=0.6]
\begin{axis} 
[x=1cm,y=1cm,xmin=0,xmax=9,ymin=0,
axis x line=none,
axis y line=none
]
\addplot [mark=none,thick]file {grid_hill.dat};
\end{axis}

\node [above] at (4.5,3.035) {$L=9H$};
\node [right] at (9.0,3.035) {$y=3.035H$};
\node [right] at (9.0,1) {$y=H$};


\draw [myarrow-myarrow] (-1,1) -- (-1,0) -- (0,0);
\node [below right] at (0,0) {$x$};
\node [above] at (-1,1) {$y$};

\end{tikzpicture}
\caption{The geometry of the hill.}
\label{hill-geom}
\end{figure}
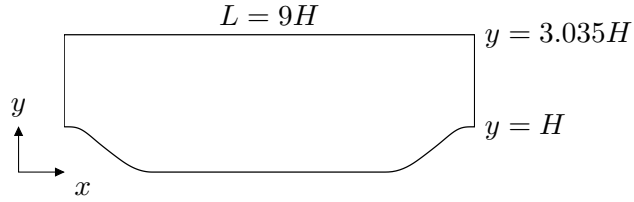

\begin{figure}
\centering
\begin{subfigure}[t]{0.5\textwidth}
\centering
\includegraphics[scale=0.28,clip=]{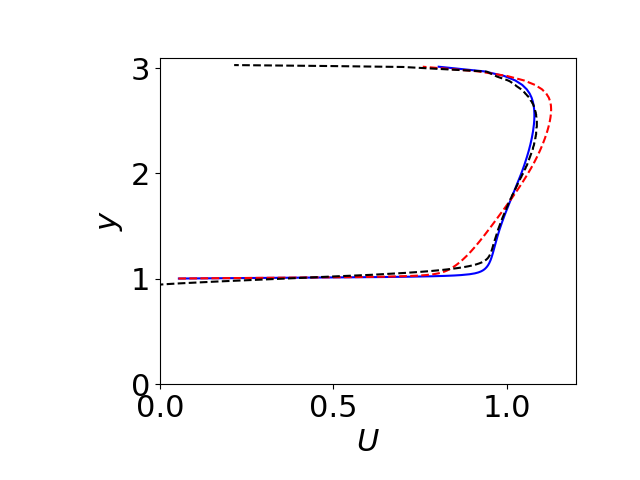}
\caption{$x=0.05$.}
\label{u05-hill}
\end{subfigure}%
\begin{subfigure}[t]{0.5\textwidth}
\centering
\includegraphics[scale=0.28,clip=]{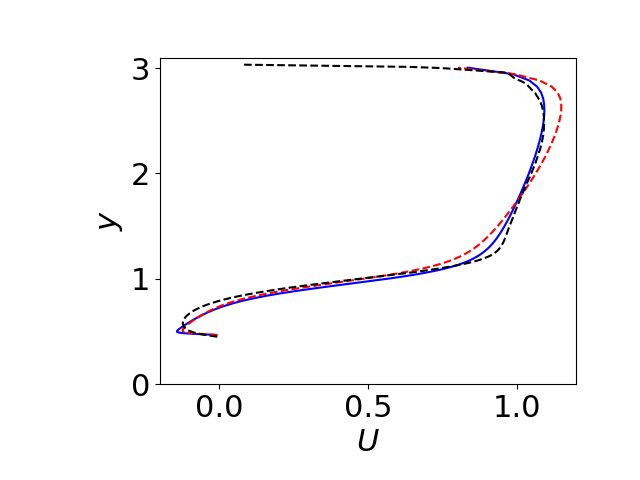}
\caption{$x=1$.}
\label{u1-hill}
\end{subfigure}
\begin{subfigure}[t]{0.5\textwidth}
\centering
\includegraphics[scale=0.28,clip=]{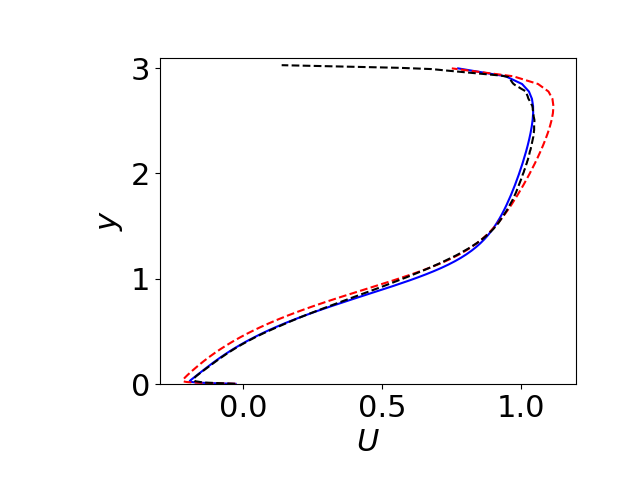}
\caption{$x=3$.}
\label{u3-hill}
\end{subfigure}%
\begin{subfigure}[t]{0.5\textwidth}
\centering
\includegraphics[scale=0.28,clip=]{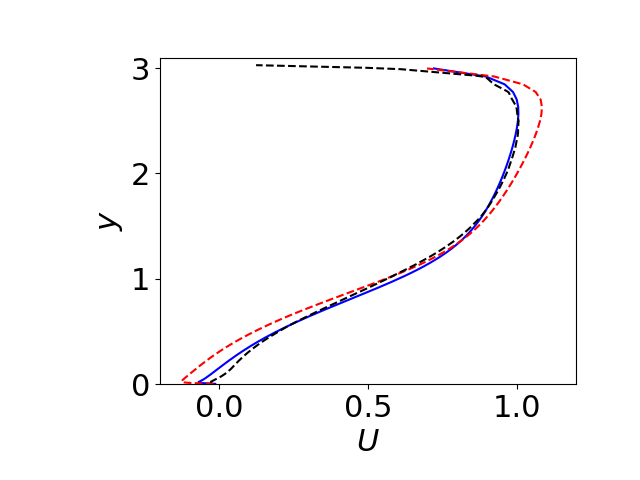}
\caption{$x=4$.}
\label{u4-hill}
\end{subfigure}
\begin{subfigure}[t]{0.5\textwidth}
\centering
\includegraphics[scale=0.28,clip=]{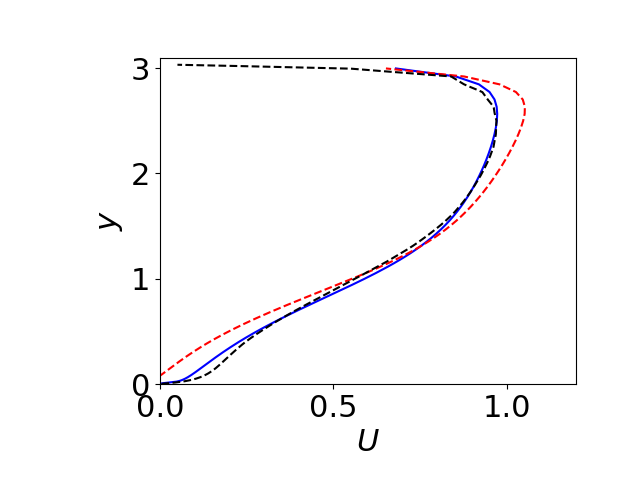}
\caption{$x=5$.}
\label{u5-hill}
\end{subfigure}%
\begin{subfigure}[t]{0.5\textwidth}
\centering
\includegraphics[scale=0.28,clip=]{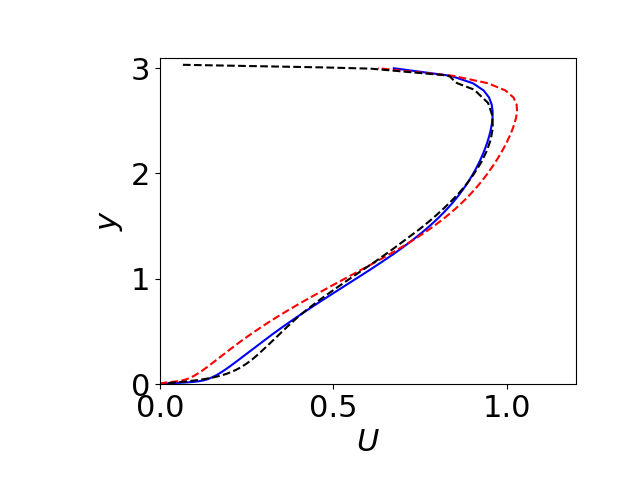}
\caption{$x=7$.}
\label{u7-hill}
\end{subfigure}
\caption{Hill flow. Velocity. \solidline: $k-\omega$-PINN-NN; \dashedline: standard $k-\omega$; 
\dashedblackline: DNS~\cite{froehlich:mellen:rodi:temmerman:leschziner:05}}
\label{hill-u}
\end{figure}

\begin{figure}
\centering
\begin{subfigure}[t]{0.5\textwidth}
\centering
\includegraphics[scale=0.28,clip=]{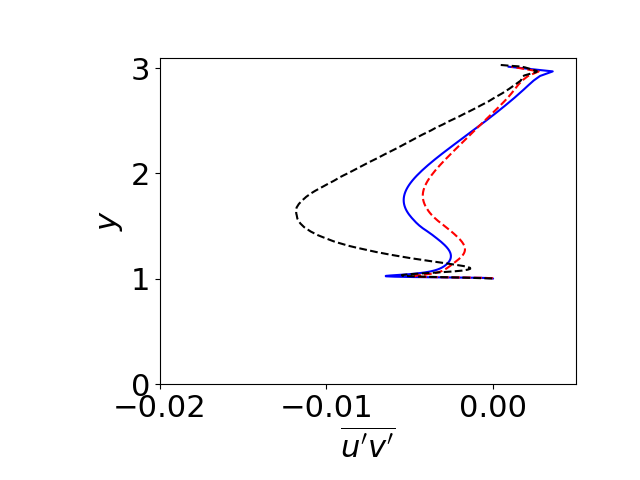}
\caption{$x=0.05$.}
\label{uv05-hill}
\end{subfigure}%
\begin{subfigure}[t]{0.5\textwidth}
\centering
\includegraphics[scale=0.28,clip=]{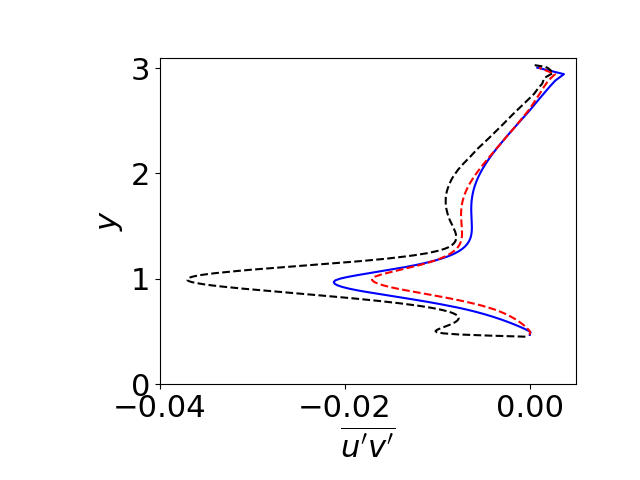}
\caption{$x=1$.}
\label{uv1-hill}
\end{subfigure}
\begin{subfigure}[t]{0.5\textwidth}
\centering
\includegraphics[scale=0.28,clip=]{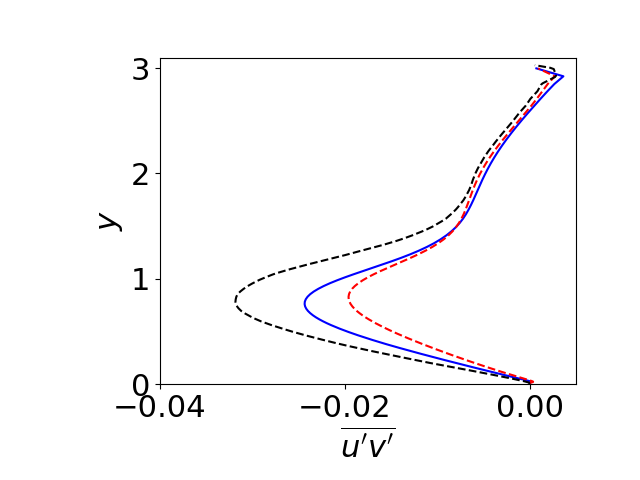}
\caption{$x=3$.}
\label{uv3-hill}
\end{subfigure}%
\begin{subfigure}[t]{0.5\textwidth}
\centering
\includegraphics[scale=0.28,clip=]{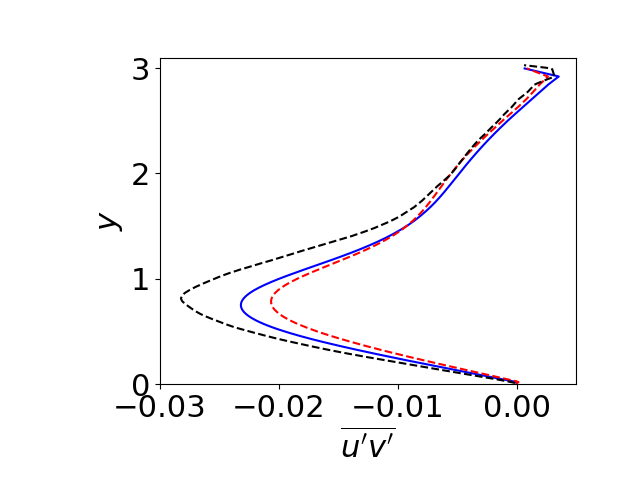}
\caption{$x=4$.}
\label{uv4-hill}
\end{subfigure}
\begin{subfigure}[t]{0.5\textwidth}
\centering
\includegraphics[scale=0.28,clip=]{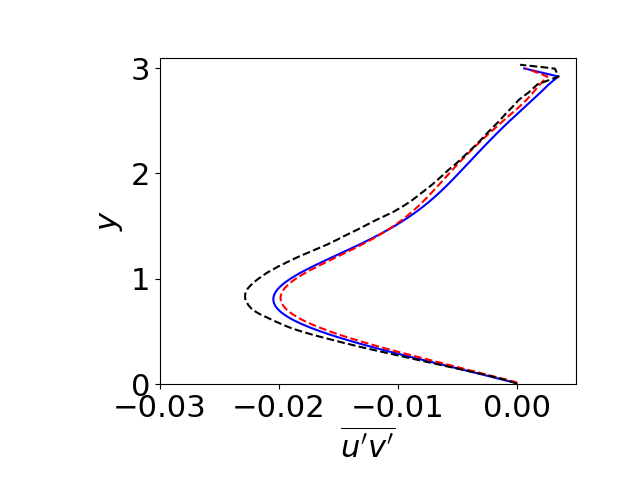}
\caption{$x=5$.}
\label{uv5-hill}
\end{subfigure}%
\begin{subfigure}[t]{0.5\textwidth}
\centering
\includegraphics[scale=0.28,clip=]{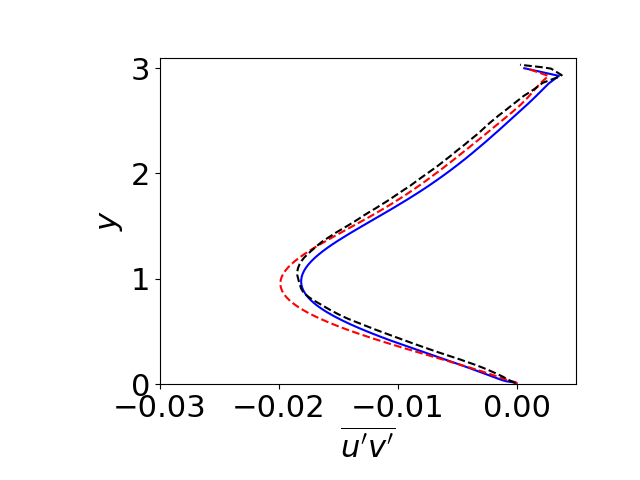}
\caption{$x=7$.}
\label{uv7-hill}
\end{subfigure}
\caption{Hill flow. Turbulent shear stresses. \solidline: $k-\omega$-PINN-NN; \dashedline: standard $k-\omega$; 
\dashedblackline: DNS~\cite{froehlich:mellen:rodi:temmerman:leschziner:05}}
\label{hill-uv}
\end{figure}

\begin{figure}
\centering
\begin{subfigure}[t]{0.5\textwidth}
\centering
\includegraphics[scale=0.28,clip=]{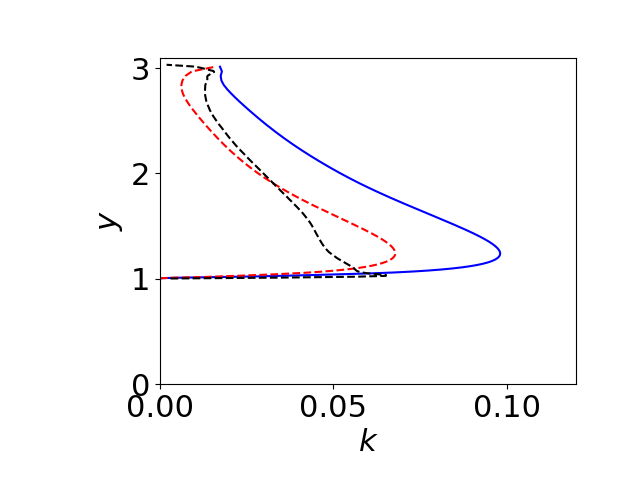}
\caption{$x=0.05$.}
\label{k05-hill}
\end{subfigure}%
\begin{subfigure}[t]{0.5\textwidth}
\centering
\includegraphics[scale=0.28,clip=]{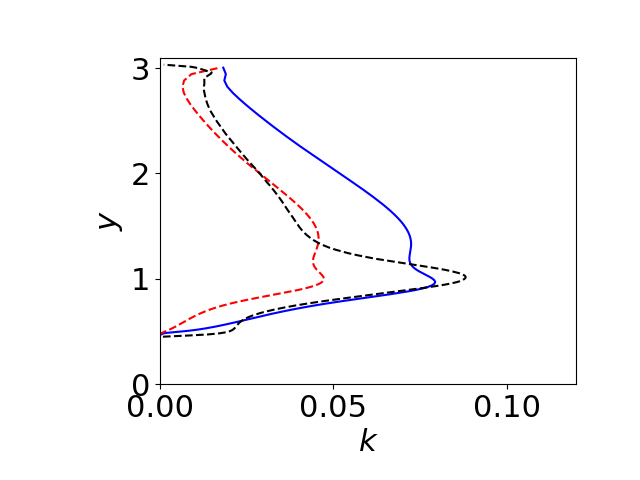}
\caption{$x=1$.}
\label{k1-hill}
\end{subfigure}
\begin{subfigure}[t]{0.5\textwidth}
\centering
\includegraphics[scale=0.28,clip=]{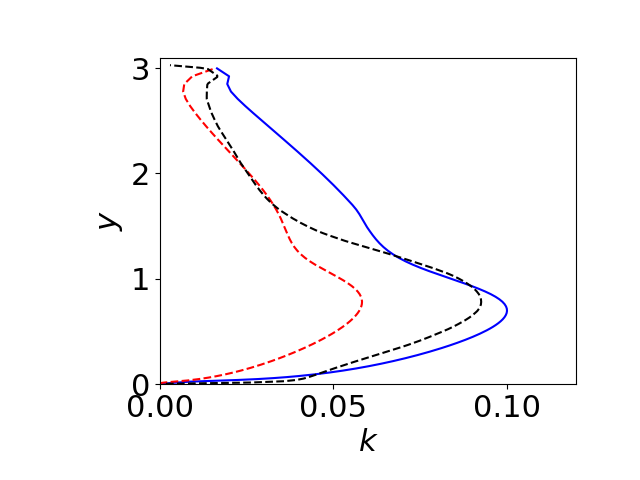}
\caption{$x=3$.}
\label{k3-hill}
\end{subfigure}%
\begin{subfigure}[t]{0.5\textwidth}
\centering
\includegraphics[scale=0.28,clip=]{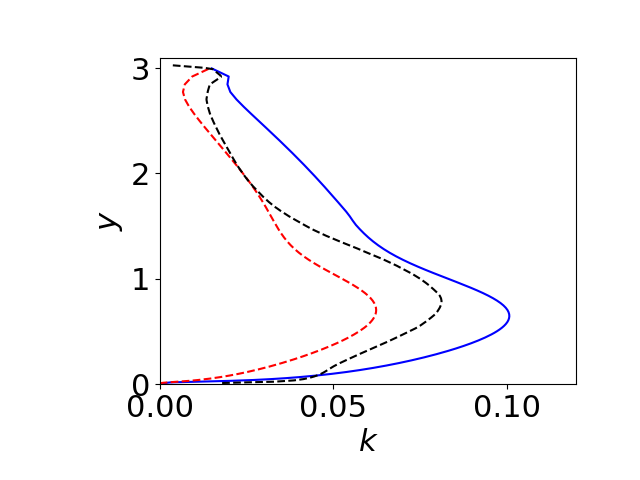}
\caption{$x=4$.}
\label{k4-hill}
\end{subfigure}
\begin{subfigure}[t]{0.5\textwidth}
\centering
\includegraphics[scale=0.28,clip=]{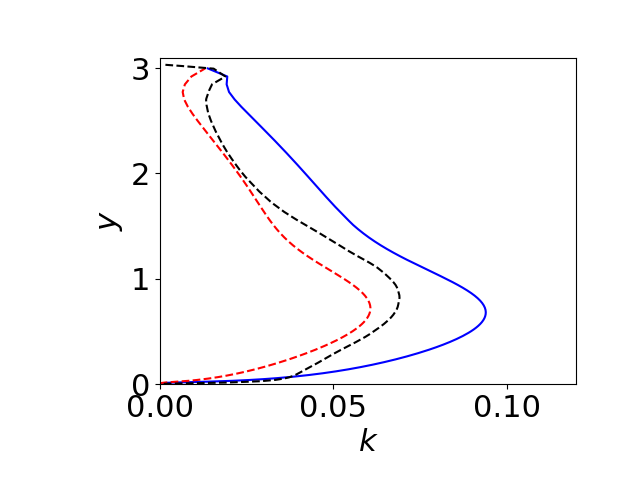}
\caption{$x=5$.}
\label{k5-hill}
\end{subfigure}%
\begin{subfigure}[t]{0.5\textwidth}
\centering
\includegraphics[scale=0.28,clip=]{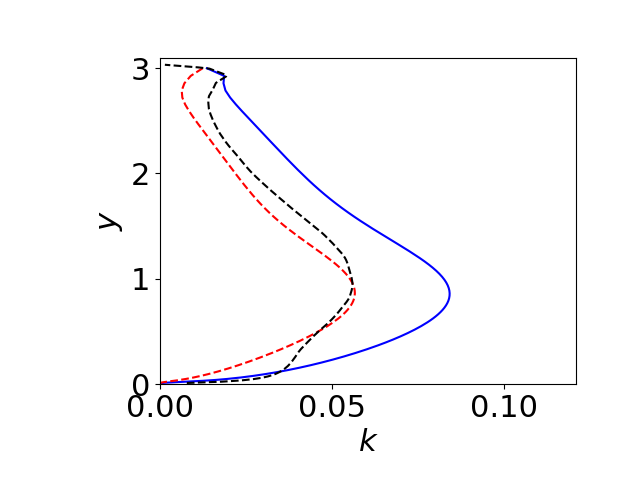}
\caption{$x=7$.}
\label{k7-hill}
\end{subfigure}
\caption{Hill flow. Turbulent kinetic energy. \solidline: $k-\omega$-PINN-NN; \dashedline: standard $k-\omega$; 
\dashedblackline: DNS~\cite{froehlich:mellen:rodi:temmerman:leschziner:05}}
\label{hill-k}
\end{figure}

\begin{figure}
\centering
\begin{subfigure}[t]{0.5\textwidth}
\centering
\includegraphics[scale=0.28,clip=]{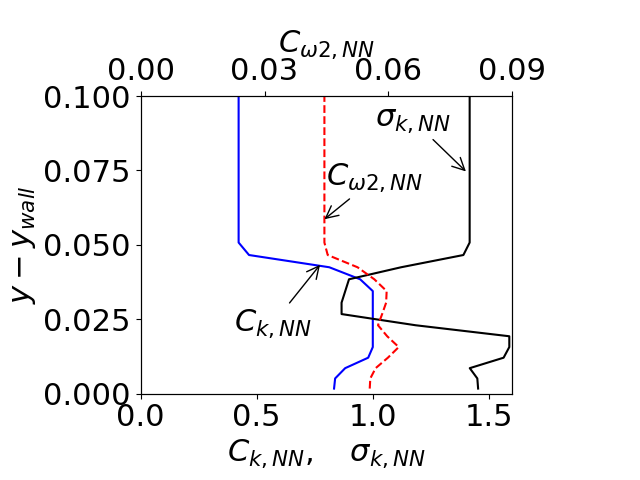}
\caption{$x=0.05$.}
\label{prand_k05-hill}
\end{subfigure}%
\begin{subfigure}[t]{0.5\textwidth}
\centering
\includegraphics[scale=0.28,clip=]{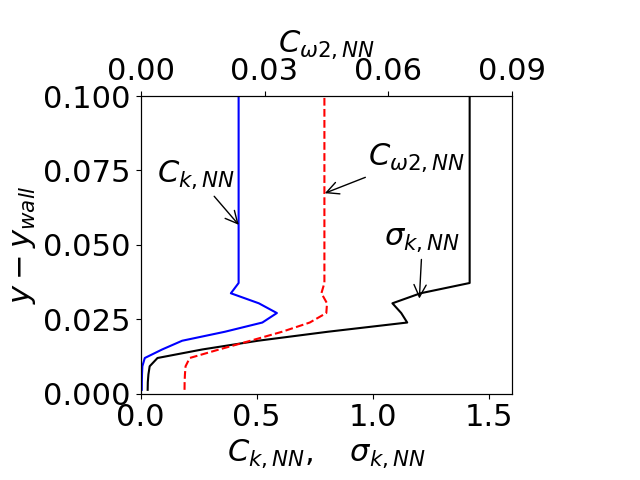}
\caption{$x=1$.}
\label{prand_k1-hill}
\end{subfigure}
\begin{subfigure}[t]{0.5\textwidth}
\centering
\includegraphics[scale=0.28,clip=]{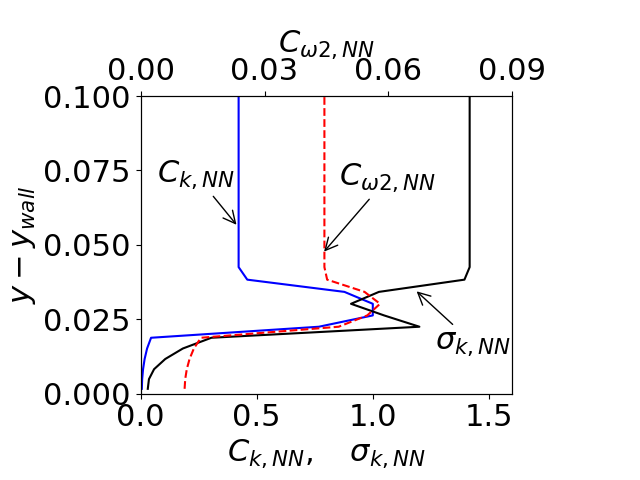}
\caption{$x=3$.}
\label{prand_k3-hill}
\end{subfigure}%
\begin{subfigure}[t]{0.5\textwidth}
\centering
\includegraphics[scale=0.28,clip=]{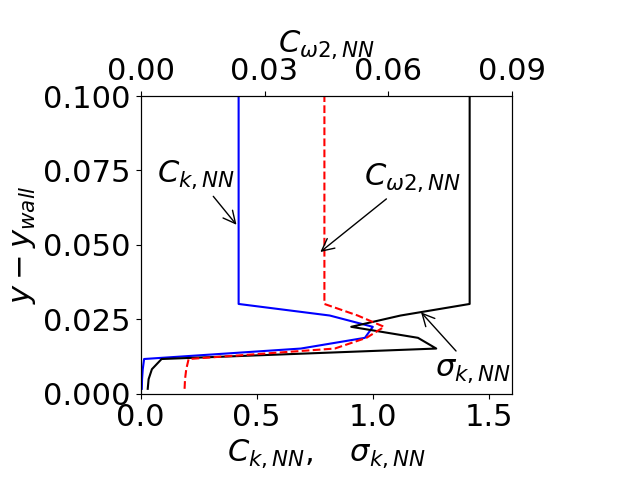}
\caption{$x=4$.}
\label{prand_k4-hill}
\end{subfigure}
\begin{subfigure}[t]{0.5\textwidth}
\centering
\includegraphics[scale=0.28,clip=]{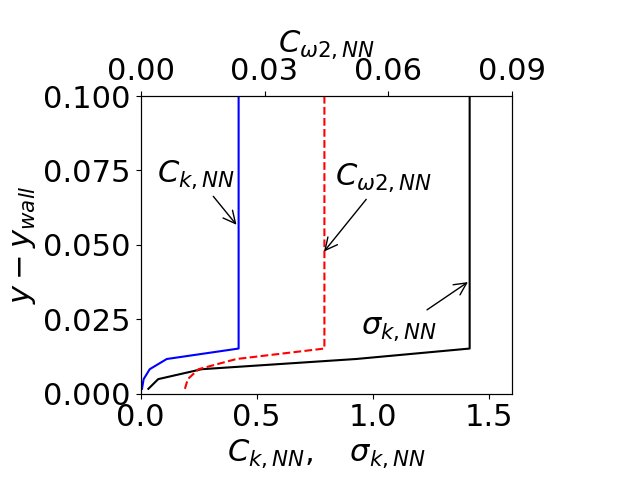}
\caption{$x=5$.}
\label{prand_k5-hill}
\end{subfigure}%
\begin{subfigure}[t]{0.5\textwidth}
\centering
\includegraphics[scale=0.28,clip=]{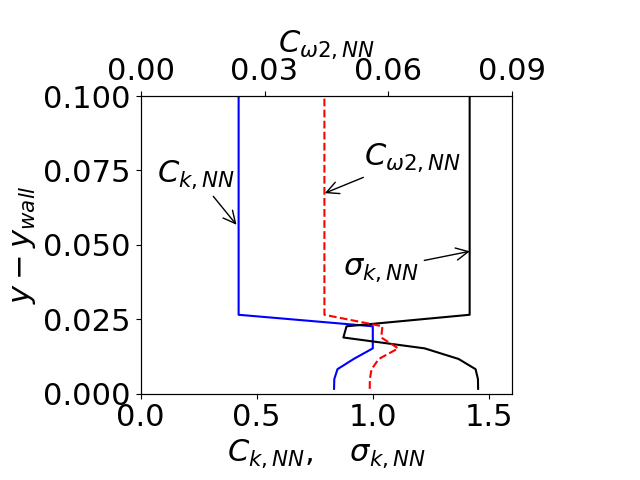}
\caption{$x=7$.}
\label{prand_k7-hill}
\end{subfigure}
\caption{Hill flow. Turbulent Prandtl number, $\sigma_{k,NN}$, $C_{k,NN}$ and $C_{\omega 2, NN}$ predicted
by the NN model. Zoomed-in view.}
\label{hill-prand_k}
\end{figure}

\begin{figure}
\centering\captionsetup[subfigure]{aboveskip=-20pt,belowskip=-20pt,justification=centering}
\centering
\begin{subfigure}[t]{0.5\textwidth}
\centering
\centering
\includegraphics[scale=0.48,clip=]{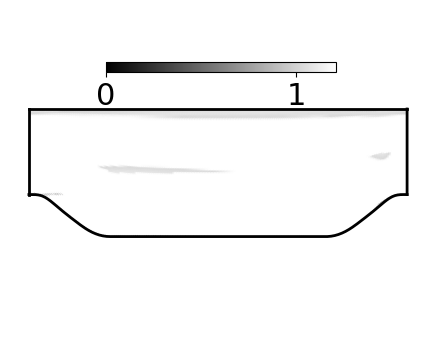}
\caption{$\sigma_{k,NN}$. Maximum value of $\sigma_{k,NN}$ in color bar is $1.21$.}
\label{prand_k-hill}
\end{subfigure}%
\begin{subfigure}[t]{0.5\textwidth}
\centering
\includegraphics[scale=0.48,clip=]{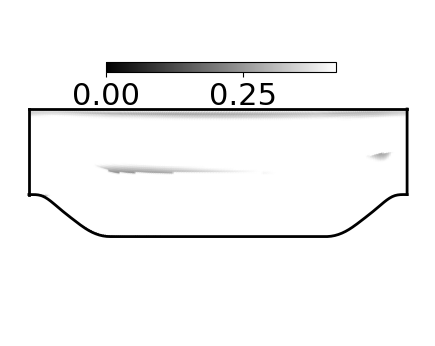}
\caption{$c_{k,NN}$. Maximum value of $C_{k,NN}$ in color bar is $0.41$.}
\label{c_k-hill}
\end{subfigure}
\begin{subfigure}[t]{0.5\textwidth}
\centering
\includegraphics[scale=0.48,clip=]{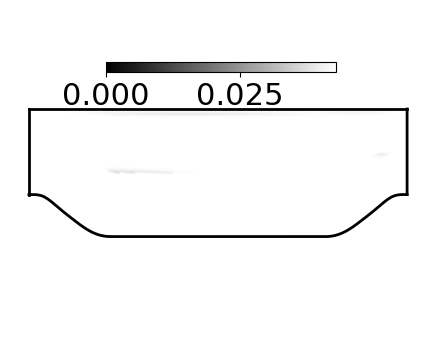}
\caption{$c_{\omega 2,NN}$.  Maximum value of $C_{\omega 2,NN}$ in color bar is $0.043$.}
\label{c_omega_2-hill}
\end{subfigure}%
\begin{subfigure}[t]{0.5\textwidth}
\centering
\includegraphics[scale=0.48,clip=]{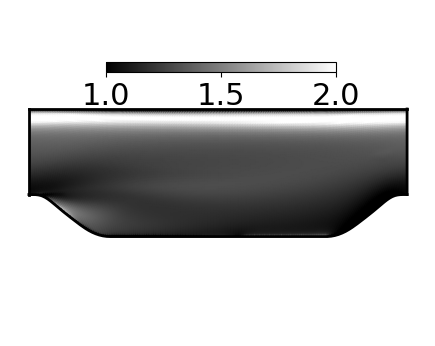}
\caption{Ratio of $\nu_{tot,NN}$ to $\nu_{tot,k-\omega}$.  Maximum value of the ratio in color bar is $2.0$.}
\label{vist-ratio}
\end{subfigure}%
\caption{Hill flow. Gray-scale plots of $\sigma_{k,NN}$,  $c_{k,NN}$,  $c_{\omega 2,NN}$ and
ratio of total  viscosities.} 
\label{hill-gray}
\end{figure}

\section{Results}
\label{results}

NN models for the turbulent Prandtl number, $\sigma_{k,NN}$,  the coefficients $C_{k,NN}$ and
 $C_{\omega 2,NN}$ were developed in the previous section.
They are in this section used in Eq.~\ref{kom} when solving the $k$ and $\omega$ equations.
 The new model is called the $k-\omega$-PINN-NN model. 

The NN models are incorporated in the CFD code \pycalcr\ as follows (the procedure is identical in \pycalc).

\begin{enumerate}
\item At the first iteration, load the NN model into \pycalcr\
\item Solve the $\vb_1$ and  $\vb_2$ equations (Eq.~\ref{RANS}) and the continuity equation (Eq.~\ref{contRANS}, 
 i.e. the pressure correction equation).
\item Call the NN models to compute  $\sigma_{k,NN}$,  $C_{k,NN}$ and $C_{\omega 2,NN}$ 
\item Solve $k$ and $\omega$  equations.
\item Compute $\nu_t = k/\omega$
\item End of global iteration. Repeat from Step 2 until convergence (1000s of iterations)
\end{enumerate}

The new model is validated in three types of flows: fully-developed channel flow
 ($Re_\tau = 550$, $2\, 000$, $5\, 200$ and  $Re_\tau = 10\, 000$), flat-plate boundary layer flow and flow over a hill.
The boundary conditions at the walls are  $\bar{u} = \bar{v}  = k=0$; $\omega$ is at the center of the wall-adjacent cells 
fixed according to Eq.~\ref{omega_wallbc}.

\subsection{Fully-developed channel flow.}

Fully-developed, i.e. one-dimensional,  channel flows are simulated 
using \pycalcr. The $\bar{u}$, $k$ and $\omega$ equations are solved. 
The domain covers only the lower half of the channel. The wall-normal gradient is set to zero for all variables at the 
upper boundary. The flow is driven by a prescribed pressure gradient (the first term on the right-hand side of Eq.~\ref{RANS}).
Details of the grids are provided in Table~\ref{tabell-channel}.
All quantities are made non-dimensional with half-channel width, $\delta$,  and friction velocity, $u_\tau$. 
Quantities with superscript
$+$ are made  non-dimensional with kinematic viscosity, $\nu$, and  $u_\tau$.

The predicted velocity, turbulent kinetic energy
and turbulent viscosity using the Wilcox $k-\omega$ model and the $k-\omega$-PINN-NN model are compared with DNS  in 
Figs.~\ref{channel-550}, \ref{channel-2000}, \ref{channel-5200} and \ref{channel-10000}. 
The velocity profiles are well predicted with both models except at the lowest Reynolds number, Fig.~\ref{vel-550},
for which the $k-\omega$-PINN-NN model gives slightly too large values. The turbulent kinetic energy is much better 
predicted with the  
$k-\omega$-PINN-NN model. It can be seen that the turbulent viscosities predicted by the two turbulence models are similar
for all Reynolds numbers.
It may be recalled that this was indeed a requirement when developing the $k-\omega$-PINN-NN 
model in the previous section, see Eq.~\ref{nut}.
They are \emph{very} similar in
the inner part of the boundary layer ($y \lesssim 0.05$, $y \lesssim 0.1$, $y \lesssim 0.18$ and 
$y \lesssim 0.29$ for $Re_\tau = 550$, $2\, 000$,
$5\, 200$ and $Re_\tau = 10\, 000$, respectively). 

Figures~\ref{prand_k-550}, \ref{prand_k-2000}, \ref{prand_k-5200} and \ref{prand_k-10000} present 
$\sigma_{k,NN}$, $C_{k,NN}$ and  $C_{\omega 2,NN}$ predicted by the NN model. They are very similar for all Reynolds numbers.
The reason is that they were made functions of the input parameters
$\tau_{tot}/u_\tau^2$ and $\nu_t/(y u_\tau)$; the former is independent of Reynolds number (see Eq.~\ref{y-1}) and the latter is only 
weekly dependent on Reynolds number (see Figs.~\ref{vist-550}, \ref{vist-2000}, \ref{vist-5200}, \ref{vist-10000}).
The reason for the over-predicted velocity profile in Fig.~\ref{vel-550} is probably that the input
parameters should be modified at this low Reynolds number.

From the zoomed-in views in Figs.~\ref{prand_k-550-zoom}, \ref{prand_k-2000-zoom}, \ref{prand_k-5200-zoom} and
\ref{prand_k-10000-zoom} it is clearly seen that the extent of the near-wall region in 
which $\sigma_{k,NN}$, $C_{k,NN}$ and  $C_{\omega 2,NN}$ diminishes as the Reynolds number increases. 
This mimics the effect of Reynolds number on the input parameter $\nu_t/(u_\tau y)$, see
Fig.~\ref{nut-over-y}.

It should be mentioned that $\sigma_{k,NN}$, $C_{k,NN}$ and  $C_{\omega 2,NN}$  exhibit slow oscillations with respect to iteration number.
Figure~\ref{c_k-iter-10000} shows how $\sigma_{k,NN}$ oscillates with respect to iteration number; the 
oscillation period is approximately $3\,000$ iterations. This makes it difficult to converge the $\omega$ equation properly.
The remedy is to introduce an
 exponentially weighted averaging operator acting on $\sigma_{k,NN}$, $C_{k,NN}$ and  $C_{\omega 2,NN}$ over $m$ iterations.
 The average of $\sigma_{k,NN}$, for example,  is defined as
\citep{xiao:12,davidson:21a}
\begin{equation}
\label{int}
\langle  \sigma_{k,NN}\rangle^{n}_T = 
a \langle \sigma_{k,NN} \rangle^{n-1}_T + (1-a) \sigma_{k,NN}^{n}, \quad a = \exp(-1/m)
\end{equation}
$\langle \cdot \rangle_T$
indicates a time average over time, $T$, i.e.
\begin{equation*}
\begin{split}
\langle \phi(t) \rangle_T &= \frac{1}{T}\int_{-\infty}^t \phi(\tau) \exp(-(t-\tau)/T)d\tau  \Rightarrow \\
\langle \phi \rangle_T^{n}  &\equiv \langle \phi \rangle_T = a \langle \phi \rangle_T^{n-1} + (1-a) \phi^n,
\end{split}
\end{equation*}
where $a=1/(1+\Delta t/T)$ and $n$ denotes the timestep number.

Taking guidance from Fig.~\ref{c_k-iter-10000} we set $m=3\,000$.  For the other three Reynolds numbers $m=500$.
The problem of slow variations of $\sigma_{k,NN}$, $C_{k,NN}$ and  $C_{\omega 2,NN}$ does not occur
in the flat-plat boundary layer or the hill flow. This issue is probably related to the strong elliptic character
of fully-developed channel flow,  i.e. the transport takes place only through diffusion.

\subsection{Flat-plat boundary layer flow.}

This flow is also  simulated using \pycalcr.
The Reynolds number at the inlet is $Re_{\theta}=2\, 550$.
The mean profiles are taken
from a pre-cursor 2D RANS simulation.
The grid has $150 \times 90$ cells ($x$, $y$).
The domain size is
$92 \delta_{in} \times 20 \delta_{in}$ where $\delta_{in}$ denotes the boundary-layer thickness at the inlet.
The grid is stretched in the wall-normal direction by $10\%$ up to $y=6.7\delta_{in}$  
where the cell size is $\Delta_y  =   0.6\delta_{in}$; it is stretched by $1\%$ in the streamwise direction.

Figure~\ref{bound} presents comparisons of predictions by the $k-\omega$-PINN-NN model and the Wilcox $k-\omega$ with DNS.
It can be seen that the skin friction and the velocity profile are very good predicted by both turbulence models.
The peak of the shear stress is also well predicted by both models but the magnitude of the shear stress is slightly too 
small in the outer region.
However, the predicted peak of the turbulent kinetic energy by the $k-\omega$-PINN-NN model is too large but its form
is much better than that predicted with the Wilcox $k-\omega$ model. The predicted $\sigma_{k,NN}$, $C_{k,NN}$ and  $C_{\omega 2,NN}$
(see Fig.~\ref{prand_k-bound}) are similar to those for the channel flows.

\subsection{Hill flow.}

The final test case is the flow over a two-dimensional periodic hill, see Fig.~\ref{hill-geom}. 
This flow is simulated using \pycalc.
The size of the domain is  $9H \times 3.035H \times H$
in the streamwise ($x$) and  the wall-normal ($y$) direction ($z$), respectively.
 The Reynolds number is  $Re=10\, 600$ based on the
hill height, $H$,  and the bulk velocity $U_b$ at the top of the hill. 
Periodic boundary conditions
are used in the $x$ direction. 
 Slip conditions are prescribed in the $z$ direction.

Unsteady simulations are carried out marching to steady-state conditions. The time step is set 
to $0.01$ which gives a maximum CFL of $0.4$. 
Two global iterations (solving $\bar{u}$, $\bar{v}$, $\bar{w}$, $k$ and $\omega$) are carried out each time step.
The flow is solved over $40\, 000$ time steps which corresponds to approximately $44$ through-flows; 
the flow variables are time-averaged over the last $100$ time steps.

A wall function is used at the upper  wall for the hill flow. It is  based on  \reich
\begin{eqnarray}
\label{reich}
\frac{\bar{u}_P}{u_\tau} \equiv U_P^+ =  \nonumber  \\
\frac{1}{\kappa} \ln(1-0.4y_P^+)+7.8\left[ 1-\exp\left(-y_P^+/11\right) - 
(y_P^+/11)\exp\left(-y_P^+/3\right)\right]
\end{eqnarray}
The friction velocity is then obtained by re-arranging Eq.~\ref{reich} and solving the implicit equation
\begin{eqnarray}
\label{ustar-reich}
u_\tau  - \bar{u}_P\left\{ 
\ln(1-0.4y_P^+)/\kappa+\right . \nonumber  \\
\left . 7.8\left[ 1-\exp\left(-y_P^+/11\right) - 
(y_P^+/11)\exp\left(-y_P^+/3\right)\right]\right\}^{-1} = 0
\end{eqnarray}
using the Newton-Raphson method \verb+scipy.optimize.newton+ in Python.
$\bar{u}_P$ and $y_P^+ = u_\tau y/\nu$ denote the
wall-parallel velocity and non-dimensional wall distance, respectively,
at the first wall-adjacent cells. 
The boundary condition for the wall-parallel velocity is the shear stress boundary condition, $\rho u_\tau^2$.
The turbulent kinetic energy, $k$,
 and its specific dissipation, $\omega$,  are set at the wall-adjacent cells centers as
\begin{equation*}
\begin{split}
C_\mu^{-1/2} u_\tau^2:&\;  k \; \rm{equation}\\
\frac{u_\tau}{\kappa y}:&\;  \omega\;  \rm{equation}
\end{split}
\end{equation*}
where $\kappa=0.4$.

The bulk velocity is kept constant by adjusting $\beta$ in Eq.~\ref{mom1} at each time step and iteration.
The coefficient in Eq.~\ref{mom1} is computed as
\begin{equation}
\label{beta}
\beta^n = \beta^{n-1} + 0.001(U_{T} - 2U_b^{n-1} + U_b^{n})
\end{equation}
where $U_{T}=1$ is the target bulk velocity at the hill crest and $n$ is the time step number.

The predictions are compared with DNS~\cite{froehlich:mellen:rodi:temmerman:leschziner:05}.
The grid is the same as in the DNS in the $x-y$ plane,  i.e. $  196\times 128$ ($x\times y$) cells. 
Two cells are used in the $z$ direction.
All quantities are normalized using the height of the hill $(H$) and the bulk velocity ($U_b$) at the crest of the hill.

Figures~\ref{hill-u} compare the predicted velocity profiles with DNS and as can be seen the agreement with the $k-\omega$-PINN-NN
model is much better than with the standard $k-\omega$ model. 
The shear stresses are slightly better predicted by the  $k-\omega$-PINN-NN
model model than by the standard $k-\omega$ model 
(which is logical as the velocity profiles are better predicted by the former model). However,
it is seen that both models predict much too small magnitude of the  Reynolds shear stresses in the outer region
near and downstream of the crest of the hill (Figs.~\ref{uv05-hill} and \ref{uv1-hill}). 
The reason may be that there is a large-scale flapping motion near the crest in the DNS that is not captured in steady RANS.

Figures~\ref{hill-k} show  the predicted turbulent kinetic energy 
 profiles.  As can be seen the predicted $k$ with the $k-\omega$-PINN-NN model is in general too large; 
the agreement for the standard  $k-\omega$ model is actually better for $x\ge 5$. 
It should be mentioned that also the steady full Reynolds-stress closures gives inaccurate
turbulent kinetic energy~\citep{jakirlic:15} (much too small values).

Figure~\ref{hill-prand_k} presents a zoomed-in view of the predicted $\sigma_{k,NN}$, $C_{k,NN}$ and  $C_{\omega 2,NN}$
by the NN model.  It can be seen that
for $y-y_{wall} \gtrsim 0.03$  they take a constant value at all six $x$ locations.
Gray-scale plots in Figs.~\ref{prand_k-hill}, \ref{c_k-hill} and \ref{c_omega_2-hill} confirm
 that they are constant in almost the entire domain.
 Away from the wall, $\sigma_{k,NN}$, $C_{k,NN}$ and  $C_{\omega 2,NN}$  take the values  $1.42$, $0.42$ and 
 $0.043$, respectively (in $88\%$ of the cells in the entire domain we find that $1.42 \le \sigma_{k,NN} \le 1.43$, 
$0.42\le C_{k,NN} \le 0.42$ and $0.043 \le C_{\omega 2,NN} \le 0.044$). The reason is that the input 
parameters to the NN model are limited to the minimum and maximum values during the training process of the NN model,
see lines 42-49 in  Listing~\ref{PINN_NN} in the Appendix.
The maximum values of input parameters
 $\tau_{tot}/u_\tau^2$ and $\nu_t/(y u_\tau)$ in the NN model are $0.995$ and  $0.37$, respectively.
The question now arises: what happens if we use these  constant values
$(\sigma_{k,NN}, C_{k,NN}, C_{\omega 2,NN} = 1.42, 0.42, 0.043$) in the 
entire domain?
The answer is that the predicted profiles in Figs.~\ref{hill-u} -- \ref{hill-k} are very similar (not shown).
However, the $k-\omega$-PINN-NN  model with these constants fails in the channel flows and the flat-plate boundary layer flow;
for example, the centerline velocity in channel flow at $Re_\tau = 10\, 000$ and the skin friction 
in the flat-plate boundary layer are $10\%$ too low and $17\%$ too high at $Re_\theta = 4\, 500$, respectively (not shown).

Figure~\ref{vist-ratio} presents the ratio of 
$\nu_{tot,NN}$ to $\nu_{tot,k-\omega}$. The peak value is $2.3$ and it is smaller than one in a few points near the lower wall.
Integrated over the entire domain, the ratio is $1.26$.

\begin{figure}
\centering
\includegraphics[scale=0.28,clip=]{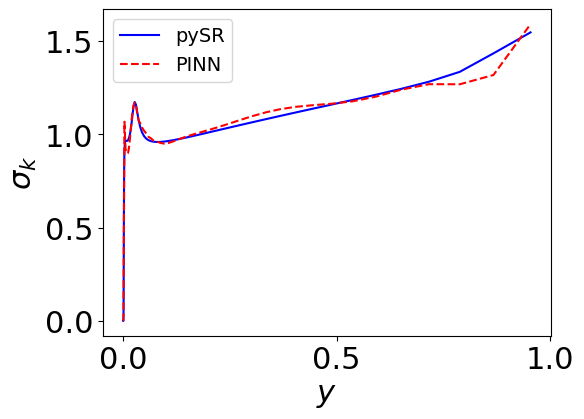}
\caption{$\sigma_k$ predicted with pySR.}
\label{prand_symb}
\end{figure}

\section{Conclusions}

The present work proposes a methodology for using PINN and NN for improving turbulence models.
A new $k-\omega$ turbulence model, $k-\omega$-PINN-NN,  is presented. 
The $k$ equation is turned into an ODE for the turbulent viscosity, $\nu_{t,PINN}$,  by  taking $k$, $P^k$ and $\e$ from DNS.
Then the turbulent Prandtl number in the $k$ equation is given by $\sigma_{k,PINN} = \nu_t/ \nu_{t,PINN}$. In order to keep
the turbulent viscosity, $\nu_t$, the same as in the standard $k-\omega$ model, two new functions are introduced:
one, $C_{k,PINN}$, in front of the dissipation term in the $k$ equation
and another, $C_{\omega 2,PINN}$, in front of the destruction term in the $\omega$
equation. They are both obtained from the $k$ and $\omega$ equations using DNS data.

Next, three NN models were created for predicting the turbulent Prandtl number, $\sigma_{k,NN}$, $C_{k,NN}$ and 
 $C_{\omega 2,NN}$. 
Two input parameters,  $\tau_{tot}/u_\tau^2$ and $\nu_t/(y u_\tau)$, were used.
The difference between $\sigma_{k,PINN}$ and $\sigma_{k,NN}$ is that the former is obtained from $\nu_{t,PINN}$ (which is 
predicted by PINN) whereas the latter is predicted by the NN model using $\sigma_{k,PINN}$ as the target.
The differences between  $C_{k,PINN}$ (or $C_{\omega 2,PINN}$) and $C_{k,NN}$ (or $C_{\omega 2,NN}$) is similar: the former
(subscript $PINN$)
are obtained via PINN and DNS data whereas the latter (subscript $NN$) are predicted by the NN models using the former as targets.

The new $k-\omega$-PINN-NN is shown to predict good skin friction, velocity and $k$ profiles in flat-plate boundary layer flow
and fully-developed channel flow. It is also found to give very good results in the periodic hill flow.
For the hill flow the NN models
predict constant values of $\sigma_{k,NN}=1.42$, $C_{k,NN}=0.42$ and  $C_{\omega 2,NN}=0.043$ at $y-y_{wall} \gtrsim 0.03$.
The reason is that the input parameters in the NN model are not allowed to exceed the values that were using when 
training the NN models. An additional simulation of the hill flow was carried out using these constant values for 
$\sigma_{k,NN}$, $C_{k,NN}$ and  $C_{\omega 2,NN}$. The predicted results were virtually the same. But when these constant
values were used for the channel flow and flat-plate boundary layer, the agreement with DNS was very poor. 

One way to further 
develop the $k-\omega$-PINN-NN  model could be to replace the NN models with symbolic regression models. For example,
the turbulent Prandtl number,  $\sigma_{k,NN}$, can be replaced with  $\sigma_{k,pySR}$ which is 
an algebraic equation.
Figure~\ref{prand_symb} shows
\begin{equation}
\label{pySR:eq}
\sigma_{k,pySR}  =   
  0.469 x_0 + \frac{0.574 + \frac{1}{49.3 + \left(x_{0} x_{1}^2 - 0.362\right)^{-1}}}{x_{0} + 0.246 + 0.0516 x_{1}/x_{0}}
\end{equation}
where $x_0 = \nu_t/(y u_\tau)$ and $x_1 = \tau_{tot}/u_\tau^2$. Equation~\ref{pySR:eq} was found
using Python symbolic regression (pySR). In Fig.~\ref{prand_symb} it is compared  with $\sigma_{k,PINN}$.
One  advantage of using Eq.~\ref{pySR:eq} instead of the NN model for $\sigma_{k,NN}$ is that Eq.~\ref{pySR:eq}
can conveniently be used in commercial CFD codes in which it may
be difficult to import Pytorch NN models.
 The pySR script can be downloaded~\citep{davidson-pinn-nn:25}.

\subsection*{Conflict of interest}

I confirm that there are no conflicts of interest.

\subsection*{Acknowledgement}

This work was financed by Chalmers University of Technology.


\begin{thebibliography}{19}
\providecommand{\natexlab}[1]{#1}
\providecommand{\url}[1]{\texttt{#1}}
\expandafter\ifx\csname urlstyle\endcsname\relax
  \providecommand{\doi}[1]{doi: #1}\else
  \providecommand{\doi}{doi: \begingroup \urlstyle{rm}\Url}\fi

\bibitem[Davidson(2025{\natexlab{a}})]{davidson-pinn-nn:25}
L.~Davidson.
\newblock Using {P}hysical {I}nformed {N}eural {N}etwork {(PINN)} and {N}eural
  {N}etwork {(NN)} to improve a $k-\omega$ turbulence model: Python {CFD} code
  and {PINN} script.
\newblock Division of Fluid Dynamics, Dept. of Mechanics and Maritime Sciences,
  Chalmers University of Technology{,} Gothenburg, 2025{\natexlab{a}}.
\newblock URL
  \url{https://www.cfd-sweden.se/lada/Using-Physical-Informed-Neural-Network-PINN-and-NN-improve-a-k-omega-turbulence-model.html}.

\bibitem[Yazdani and Tahani(2024)]{yazdani:24}
S.~Yazdani and M.~Tahani.
\newblock Data-driven discovery of turbulent flow equations using
  physics-informed neural networks.
\newblock \emph{Physics of Fluids}, 36\penalty0 (3):\penalty0 035107, 03 2024.
\newblock ISSN 1070-6631.
\newblock \doi{10.1063/5.0190138}.
\newblock URL \url{https://doi.org/10.1063/5.0190138}.

\bibitem[Luo et~al.(2020)Luo, Vellakal, Koric, Kindratenko, and Cui]{luo:20}
S.~Luo, M.~Vellakal, S.~Koric, V.~Kindratenko, and J.~Cui.
\newblock Parameter identification of {RANS} turbulence model using
  physics-embedded neural network.
\newblock In H.~Jagode, H.~Anzt, G.~Juckeland, and H.~Ltaief, editors,
  \emph{High Performance Computing}, pages 137--149, Cham, 2020. Springer
  International Publishing.
\newblock URL \url{https://link.springer.com/book/10.1007/978-3-030-59851-8}.

\bibitem[Thakur et~al.(2024)Thakur, Esmaili, Libring, Solorio, and
  Ardekani]{hakut:24}
S.~Thakur, E.~Esmaili, S.~Libring, L.~Solorio, and A.~M. Ardekani.
\newblock Inverse resolution of spatially varying diffusion coefficient using
  physics-informed neural networks.
\newblock \emph{Physics of Fluids}, 36\penalty0 (8):\penalty0 081915, 08 2024.
\newblock ISSN 1070-6631.
\newblock \doi{10.1063/5.0207453}.
\newblock URL \url{https://doi.org/10.1063/5.0207453}.

\bibitem[Wilcox(1988)]{wilcox:88}
D.~C. Wilcox.
\newblock Reassessment of the scale-determining equation.
\newblock \emph{{AIAA} Journal}, 26\penalty0 (11):\penalty0 1299--1310, 1988.

\bibitem[Lee and Moser(2015)]{moser:15}
M.~Lee and R.~D. Moser.
\newblock Direct numerical simulation of turbulent channel flow up to
  $\mathit{Re}_{{\it\tau}}\approx 5200$.
\newblock \emph{Journal of Fluid Mechanics}, 774:\penalty0 395--415, 2015.
\newblock \doi{10.1017/jfm.2015.268}.
\newblock URL \url{https://doi.org/10.1017/jfm.2015.268}.

\bibitem[Davidson(2021{\natexlab{a}})]{pycalc-rans-url}
L.~Davidson.
\newblock {pyCALC-RANS:} a {2D} {P}ython code for {RANS}.
\newblock Division of Fluid Dynamics, Dept. of Mechanics and Maritime Sciences,
  Chalmers University of Technology{,} Gothenburg, 2021{\natexlab{a}}.
\newblock URL \url{https://www.cfd-sweden.se/lada/pyCALC-RANS.html}.

\bibitem[{van Leer}(1979)]{vanleer:79}
B.~{van Leer}.
\newblock Towards the ultimate conservative difference scheme. v. a
  second-order sequel to godunov's method.
\newblock \emph{Journal of Computational Physics}, 32\penalty0 (1):\penalty0
  101--136, 1979.
\newblock ISSN 0021-9991.
\newblock \doi{https://doi.org/10.1016/0021-9991(79)90145-1}.
\newblock URL
  \url{https://www.sciencedirect.com/science/article/pii/0021999179901451}.

\bibitem[Rhie and Chow(1983)]{rhie:chow}
C.~M. Rhie and W.~L. Chow.
\newblock Numerical study of the turbulent flow past an airfoil with trailing
  edge separation.
\newblock \emph{{AIAA} Journal}, 21:\penalty0 1525--1532, 1983.

\bibitem[Davidson(2021{\natexlab{b}})]{pyCALC-LES}
L.~Davidson.
\newblock {pyCALC-LES:} a {P}ython code for {DNS}, {LES} and {H}ybrid
  {LES-RANS}.
\newblock Division of Fluid Dynamics, Dept. of Mechanics and Maritime Sciences,
  Chalmers University of Technology{,} Gothenburg, 2021{\natexlab{b}}.
\newblock URL
  \url{https://www.cfd-sweden.se/lada/postscript_files/py-calc-les.pdf}.

\bibitem[Davidson(2025{\natexlab{b}})]{davidson_etmm15}
L.~Davidson.
\newblock Using {P}hysical {I}nformed {N}eural {N}etwork {(PINN)} to improve a
  $k-\omega$ turbulence model.
\newblock In \emph{15th International ERCOFTAC Symposium on Engineering
  Turbulence Modelling and Measurements (ETMM15), Dubrovnik on 22-24
  September}, 2025{\natexlab{b}}.
\newblock URL
  \url{https://www.cfd-sweden.se/lada/Using-Physical-Informed-Neural-Network-PINN-improve-a-k-omega-turbulence-model.html}.

\bibitem[Davidson(2021{\natexlab{c}})]{davidson:MoF-url}
L.~Davidson.
\newblock Fluid mechanics, turbulent flow and turbulence modeling.
\newblock eBook, Division of Fluid Dynamics, Dept. of Mechanics and Maritime
  Sciences, Chalmers University of Technology{,} Gothenburg,
  2021{\natexlab{c}}.
\newblock URL
  \url{https://www.cfd-sweden.se/lada/postscript_files/solids-and-fluids_turbulent-flow_turbulence-modelling.pdf}.

\bibitem[Sillero et~al.(2014)Sillero, Jimenez, and Moser]{sillero:13}
J.A. Sillero, J.~Jimenez, and R.D. Moser.
\newblock One-point statistics for turbulent wall-bounded flows at {Reynolds}
  numbers up to $\delta^+\simeq 2000$.
\newblock \emph{Physics of Fluids}, 25\penalty0 (105102), 2014.
\newblock \doi{https://doi.org/10.1063/1.4823831}.
\newblock URL \url{https://doi.org/10.1063/1.4823831}.

\bibitem[Tucker(1998)]{tucker:98}
P.G. Tucker.
\newblock Assessment of geometric multilevel convergence robustness and a wall
  distance method for flows with multiple internal boundaries.
\newblock \emph{Applied Mathematical Modelling}, 22\penalty0 (4):\penalty0
  293--311, 1998.
\newblock ISSN 0307-904X.
\newblock \doi{https://doi.org/10.1016/S0307-904X(98)10007-0}.
\newblock URL
  \url{https://www.sciencedirect.com/science/article/pii/S0307904X98100070}.

\bibitem[Froehlich et~al.(2005)Froehlich, Mellen, Rodi, Temmerman, A., and
  Leschziner]{froehlich:mellen:rodi:temmerman:leschziner:05}
J.~Froehlich, C.~Mellen, W.~Rodi, L.~Temmerman, M.~A., and Leschziner.
\newblock Highly-resolved large eddy simulations of separated flow in a channel
  with streamwise periodic constrictions.
\newblock \emph{Journal of Fluid Mechanics}, 526:\penalty0 19--66, 2005.

\bibitem[Xiao and Jenny(2012)]{xiao:12}
H.~Xiao and P.~Jenny.
\newblock A consistent dual-mesh framework for hybrid {LES/RANS} modeling.
\newblock \emph{Journal of Computational Physics}, 231:\penalty0 1848--1865,
  2012.

\bibitem[Davidson(2021{\natexlab{d}})]{davidson:21a}
L.~Davidson.
\newblock Non-zonal detached eddy simulation coupled with a steady {RANS}
  solver in the wall region.
\newblock \emph{International Journal of Heat and Fluid Flow}, 92:\penalty0
  108880, 2021{\natexlab{d}}.
\newblock ISSN 0142-727X.
\newblock \doi{https://doi.org/10.1016/j.ijheatfluidflow.2021.108880}.
\newblock URL
  \url{https://www.sciencedirect.com/science/article/pii/S0142727X21001107}.

\bibitem[Jakirlic and Maduta(2015)]{jakirlic:15}
S.~Jakirlic and R.~Maduta.
\newblock Extending the bounds of 'steady' rans closures: Toward an
  instability-sensitive reynolds stress model.
\newblock \emph{International Journal of Heat and Fluid Flow}, 51:\penalty0
  175--194, 2015.
\newblock ISSN 0142-727X.
\newblock \doi{https://doi.org/10.1016/j.ijheatfluidflow.2014.09.003}.
\newblock URL
  \url{https://www.sciencedirect.com/science/article/pii/S0142727X14001180}.
\newblock Theme special issue celebrating the 75th birthdays of Brian Launder
  and Kemo Hanjalic.

\bibitem[Davidson(2026)]{davidson-autograd}
L.~Davidson.
\newblock Understanding autograd and neural network in pytorch.
\newblock Division of Fluid Dynamics, Dept. of Mechanic Engineering, Chalmers
  University of Technology{,} Gothenburg, 2026.
\newblock URL
  \url{https://www.cfd-sweden.se/lada/postscript_files/understanding-autograd-and-neural-network-in-pytorch.pdf}.

\end{thebibliography}

\appendix

\section{Physics informed NN (PINN)}

\label{PINN}

\begin{figure}
\centering
\begin{tikzpicture}[x=2.2cm,y=1.4cm]
  \message{^^JNeural network with arrows}
  \readlist\Nnod{1,2,1} 
 \foreachitem \N \in \Nnod{ 
    \edef\lay{\Ncnt} 
    \message{\lay,}
    \pgfmathsetmacro\prev{int(\Ncnt-1)} 
    \foreach \i [evaluate={\y=\N/2-\i; \x=\lay; \n=\nstyle;}] in {1,...,\N}{ 
      
      \node[node \n] (N\lay-\i) at (\x,\y) {$a_\i^{(\prev)}$};
      
      \ifnum\lay>1 
        \foreach \j in {1,...,\Nnod[\prev]}{ 
          \draw[connect arrow] (N\prev-\j) -- (N\lay-\i); 
        }
      \fi 
      
    }
  }

  \node[above=5,align=center,color=white] at (N1-1.90) {input\\[-0.2em]layer};
  \node[above=2,align=center,color=white] at (N3-1.90) {hidden layers};
  \node[above=8,align=center,color=white] at (N\Nnodlen-1.90) {output\\[-0.2em]layer};
\end{tikzpicture}
\caption{Schematic of a simple NN.  One hidden layer with two neurons denoted by
circles. The vectors between the neurons represent connections.}
\label{NN_0}
\end{figure}

Let's create a simple NN that  finds a damping function in, e.g.,  the $k-\e$ model, $Y\equiv f$,  as a function of 
$X\equiv y^+$, see Fig.~\ref{NN_0}.
It has one input ($X = a_1^{(0)}$), one hidden layer with  two neurons ($a_1^{(1)},a_2^{(1)}$) and one output ($y_{pred} =  a_1^{(2)}$).
The target is the correct $Y = f_{DNS}$ taken from DNS.
In Listing~\ref{NN_1_list-code} in Appendix~\ref{python-scripts}, you find the line labeled \texttt{Connection 0-1} which  connects 
$a_1^{(0)}$ and $\left( a_1^{(1)},a_2^{(1)}\right)$
and the line labeled \texttt{Connection 1-2} which connects $\left(a_1^{(1)}, a_2^{(1)}\right)$ and $a_1^{(2)}$.

Now we formulate the NN with weights, $w$, and biases, $b$ and 
add a Sigmoid activator, $s$,  to both neurons, i.e. 

\begin{eqnarray*}
\text{Neuron 1:} & a_1^{(1)} = s\left\{w_1^{(0)} a_1^{(0)} + b_1^{(0)} \right\} \\
\text{Neuron 2:} & a_2^{(1)} = s\left\{ w_2^{(0)} a_1^{(0)} + b_2^{(0)} \right\} \\
\text{Output:} & a_1^{(2)} = s\left\{w_1^{(1)} a_1^{(1)} + b_1^{(1)} \right . \\
  & \left .+  w_2^{(1)} a_2^{(1)} + b_2^{(1)} \right\}
\equiv Y  
\end{eqnarray*}
The expressions in the curly 
parenthesis of $s$ are the arguments of the actuator, e.g. $x$ in  \texttt{relu  = max(0,x)}.
The Python code is given in Listing~\ref{NN_0_list-code} in Appendix~\ref{python-scripts}.

The \texttt{loss.backward()} command computes the gradients of the loss, $L$, with respect to the weights, biases and activators,
(i.e. $\D L/\D w_1$, $\D L/\D b_1$, $\D L/\D s \dots$) in order to get new improved $w_1, b_1, w_2, \ldots$

For in-depth understanding of autograd and NN, the reader is referred to~\cite{davidson-autograd} and the references therein.

\section{Simple Python scripts}

\label{python-scripts}

\begin{lstlisting}[basicstyle=\fontsize{8}{8}\selectfont\ttfamily, % numbers=right,
caption={Simple NN. Initiation and forward step},captionpos=b,label={NN_1_list-code}]
class NN(nn.Module):
 def __init__(self):
  self.lay_1=nn.Linear(1, 2) # Connection 0-1
  self.lay_2=nn.Linear(2, 1) # Connection 1-2
 def forward(self, x):
  x = torch.nn.functional.sigmoid(self.lay_1(x)) 
  out = torch.nn.functional.sigmoid(self.lay_2(x))
\end{lstlisting}

\begin{lstlisting}[basicstyle=\fontsize{8}{8}\selectfont\ttfamily,caption={Simple NN. Prediction and backward step},captionpos=b,label={NN_0_list-code}]
#  initiate the NN model
model = NN()
# define input, X
X=np.zeros(nj,1))
X[:,0] = scaler_yplus.fit_transform(yplus)[:,0]
# define target Y 
Y = f_DNS
# Training loop
for epoch in range(max_no_epoch):
# Compute prediction and loss, L
 y_pred = model(X) #prediction
 L = loss_fn(y_pred, Y)  # L = |y_pred-Y|_2
 L.backward() 
\end{lstlisting}

\begin{lstlisting}[basicstyle=\fontsize{8}{8}\selectfont\ttfamily,caption={Python code for PINN.},captionpos=b,label={PINN_0}]
class MyNet(nn.Module):

    def __init__(self):
        super(MyNet, self).__init__()
        self.layer_1 = nn.Linear(1, 30)
        self.layer_2 = nn.Linear(30, 30)
        self.layer_3 = nn.Linear(30, 30)
        self.layer_4 = nn.Linear(30, 30)
        self.layer_5 = nn.Linear(30, 1)
        
    def forward(self, x):
        x = torch.nn.functional.sigmoid(self.layer_1(x)) 
        x = torch.nn.functional.sigmoid(self.layer_2(x))
        x = torch.nn.functional.sigmoid(self.layer_3(x))
        x = torch.nn.functional.sigmoid(self.layer_4(x))
        x = torch.nn.functional.sigmoid(self.layer_5(x))

        return x

def ODE(y, nut): 
 nut_y = grad(nut, y, torch.ones\
   (y.size()[0], 1,),create_graph=True)[0]
# Differential equation loss
 ODE_loss = (nu+nut)*k_yy + k_y*nut_y + Pk - eps
 ODE_loss = torch.sum(ODE_loss ** 2)
# b.c. loss. nut_0 = 0 at wall;  nut_1 = nut_DNS at centerline 
 BC_loss = (nut[0] - nut_0) ** 2  +  (nut[-1] - nut_1) ** 2
 return ODE_loss, BC_loss
nut = model(x)
loss_ODE, loss_bc = ODE(y,nut)
L = loss_ODE+100*loss_bc
L.backward()
\end{lstlisting}

\begin{lstlisting}[basicstyle=\fontsize{8}{8}\selectfont\ttfamily,caption={NN model in the CFD code (hill flow).},captionpos=b,label={PINN_NN}]
def modify_PINN(prand_k_ML, c_k_ML, c_omega_2_ML):
# load packages etc
.
.
.
    if itstep == 0 and iter == 0:
# load data c_k
      NN_c_k = torch.load('NN-model.pth',weights_only=False)
      scaler_vist_over_y = load('model-scaler-vist_over_yin')
      scaler_uv = load('model-scaler-uv_tot.bin')

      name='min-max.txt'
      uv_min,uv_max,vist_over_y_min,vist_over_y_max = xp.loadtxt(name)

# load NN models for prand_k, c_omega_2
.
.
.

# compute ustar, south wall
    ustar_s=(abs(u2d[:,0])*viscos/dist2d[:,0])**0.5
#make it 2D
    ustar_s=xp.repeat(ustar_s[:,None], repeats=nj, axis=1)
# compute ustar, north wall (from wall functions)
    ustar_n=cmu**0.25*k2d[:,-1]**0.5
#make it 2D
    ustar_n=xp.repeat(ustar_n[:,None], repeats=nj, axis=1)
    ywall_s=0.5*(y2d[0:-1,0]+y2d[1:,0])
    dist_s=yp2d-ywall_s
# dist2d = distance to nearest wall
    ustar = xp.where(dist_s > dist2d[:,0],ustar_n,ustar_s)

# strain-rate tensor
    s11 = dudx
    s12 = 0.5**(dudy+dvdx)
    s21 = s12
    s22 = dvdy
    ss = (2*(s11**2+s12**2+s21**2+s22**2))**0.5
    uv = vis2d*ss/ustar**2
    vist_over_y = (vis2d - viscos)/dist2d/ustar 

# limit min/max
# set limits on uv
    uv=xp.minimum(uv,uv_max)
    uv=xp.maximum(uv,uv_min)

# set limits on vist_over_y
    vist_over_y=xp.minimum(vist_over_y,vist_over_y_max)
    vist_over_y=xp.maximum(vist_over_y,vist_over_y_min)

# give input parameters to NN model
    vist_over_y = vist_over_y.reshape(-1,1)
    uv= uv.reshape(-1,1)
    X=xp.zeros((len(uv),2))
    X[:,0] = scaler_vist_over_y.transform(vist_over_y)[:,0]
    X[:,1] = scaler_uv.transform(uv)[:,0]
    X_tensor = torch.tensor(X, dtype=torch.float32)

# predict prand_k
    prand_k_pred = NN_prand_k(X_tensor)
# transform from tensor to numpy
    prand_k_ML = prand_k_pred.detach().numpy()[:,0]
    prand_k_ML = xp.reshape(prand_k_ML,(ni,nj,nk))

# set limits om prand_k_ML
    prand_k_ML=np.minimum(prand_k_ML,prand_k_ML_max)
    prand_k_ML=np.maximum(prand_k_ML,prand_k_ML_min)

# predict c_K_ML and c_omega_2_ML
.
.
.
    return prand_k_ML, c_k_ML, c_omega_2_ML

\end{lstlisting}

\end{document}